%% file: prometheus.tex
\documentclass[a4paper,fleqn]{cas-sc}
\input{packages.tex}

\input{commands.tex}

\begin{document}

\input{sections/preamble}
\input{sections/poem}
\input{sections/introduction}
\input{sections/overview}
\input{sections/examples}
\input{sections/output}
\input{sections/structure}
\input{sections/performance}

\input{sections/checks}
\input{sections/conclusions}
\input{sections/acknowledge}

\bibliographystyle{elsarticle-num}
\bibliography{prometheus}

\pagebreak
\appendix

\input{appendices/config.tex}
\input{appendices/examples.tex}
\input{appendices/earth_models.tex}

\end{document}

%% file: packages.tex
\usepackage{CJKutf8}
\usepackage{amsmath}
\usepackage{url}
\usepackage{listings}
\usepackage{subfig}
\usepackage[capitalise]{cleveref}
\usepackage{natbib} 
\setcitestyle{square,numbers,sort&compress}
\usepackage{tabularx} 
\usepackage{array}
\usepackage{fontawesome} 
\usepackage{lettrine} 
\usepackage{fancyhdr} 
\usepackage{poemscol} 
\input{Zallman.fd}

%% file: commands.tex
\newcommand{\leptoninjector}{\texttt{LeptonInjector}}
\newcommand{\prometheus}{\texttt{Prometheus}}
\newcommand{\particle}{\texttt{Particle}}
\newcommand{\detector}{\texttt{Detector}}
\newcommand{\propagatableparticle}{\texttt{PropagatableParticle}}

\definecolor{codegreen}{rgb}{0,0.6,0}
\definecolor{codegray}{rgb}{0.5,0.5,0.5}
\definecolor{codepurple}{rgb}{0.58,0,0.82}
\definecolor{backcolour}{rgb}{0.95,0.95,0.92}

\lstdefinestyle{mystyle}{
    backgroundcolor=\color{backcolour},   
    commentstyle=\color{codegray},
    keywordstyle=\color{codegreen},
    numberstyle=\tiny\color{codegray},
    stringstyle=\color{orange},
    basicstyle=\ttfamily\footnotesize,
    breakatwhitespace=false,         
    breaklines=true,                 
    captionpos=b,                    
    keepspaces=true,                 
    numbers=left,                    
    numbersep=5pt,                  
    showspaces=false,                
    showstringspaces=false,
    showtabs=false,                  
    tabsize=2
}

%% file: sections/preamble.tex
\let\WriteBookmarks\relax
\def\floatpagepagefraction{1}
\def\textpagefraction{.001}

\author[1,2]{Jeffrey Lazar}[
    orcid=0000-0003-0928-5025
]
\cormark[1]

\ead{jeffreylazar@fas.harvard.edu}

\author[3]{Stephan Meighen-Berger}[
    orcid=0000-0001-6579-2000
]
\cormark[1]
\ead{stephan.meighenberger@unimelb.edu.au}

\author[4]{Christian Haack}[
    orcid=0000-0003-3932-2448
]

\author[5]{David Kim}[
    orcid=0000-0002-6025-8316
]

\author[1]{Santiago Giner}

\author[1]{Carlos A. Arg\"{u}elles}[
    orcid=0000-0003-4186-4182
]

\affiliation[1]{
    organization={Department of Physics and Laboratory for Particle Physics and Cosmology},
    city={Cambridge},
    citysep={}, 
    postcode={02138}, 
    state={MA},
    country={United States}
}


\affiliation[2]{
    organization={Department of Physics and Wisconsin IceCube Particle Astrophysics Center University of Wisconsin–Madison},
    city={Madison},
    citysep={}, 
    postcode={53703}, 
    state={WI},
    country={United States}
}


\affiliation[2]{
    organization={School of Physics, The University of Melbourne, Melbourne},
    city={Melbourne},
    citysep={},
    postcode={3010}, 
    state={Victoria},
    country={Australia}
}


\affiliation[4]{
    organization={Physik-department, Technische Universit{\"a}t M{\"u}nchen},
    city={M{\"u}nchen},
    addressline={D-85748 Garching},
    country={Germany}
}


\affiliation[5]{
    organization={Department of Physics, Cornell University},
    city={Ithaca},
    citysep={}, 
    postcode={14853}, 
    state={NY},
    country={United States}
}


\cortext[cor1]{Corresponding authors}

\shortauthors{J. Lazar \textit{et~al.}}

\title [mode = title]{\texttt{Prometheus}: An Open-Source Neutrino Telescope Simulation {\raisebox{-2.50\depth}{{\href{https://github.com/Harvard-Neutrino/prometheus}{\huge\color{BlueViolet}\faGithub}}}}}

\shorttitle{\texttt{Prometheus}}

\begin{abstract}
Neutrino telescopes are gigaton-scale neutrino detectors comprised of individual light-detection units. 
Though constructed from simple building blocks, they have opened a new window to the Universe and are able to probe center-of-mass energies that are comparable to those of collider experiments.
\prometheus{} is a new, open-source simulation tailored for this kind of detector. 
Our package, which is written in a combination of \texttt{C++} and \texttt{Python} provides a balance of ease of use and performance and allows the user to simulate a neutrino telescope with arbitrary geometry deployed in ice or water.
\prometheus{} simulates the neutrino interactions in the volume surrounding the detector, computes the light yield of the hadronic shower and the out-going lepton, propagates the photons in the medium, and records their arrival times and position in user-defined regions.
Finally, \prometheus{} events are serialized into a \texttt{parquet} file, which is a compact and interoperational file format that allows prompt access to the events for further analysis.
\end{abstract}


\maketitle

%% file: sections/poem.tex
\begin{poem}
\begin{stanza}
\textit{Who helped me \verseline
Against the Titans' insolence?\verseline
Who rescued me from certain death,\verseline
From slavery?\verseline
Didst thou not do all this thyself,\verseline
My sacred glowing heart?\verseline
And glowedst, young and good,\verseline
Deceived with grateful thanks\verseline
To yonder slumbering one?}
\end{stanza}
Johann Wolfgang von Goethe, \textit{1749--1832}
\end{poem}
\vspace{-0.50cm}

%% file: sections/introduction.tex
\section{Introduction}
\label{sec:intro}


Neutrino telescopes~\cite{Klein:2022lrf} are gigaton-scale neutrino detectors that use naturally occurring media such as glaciers~\cite{ANITA:2016vrp,ANITA:2018sgj,IceCube-Gen2:2020qha}, the Pacific Ocean~\cite{P-ONE:2020ljt}, lakes~\cite{BAIKAL:2015hjt,Avrorin:2015wba}, seas~\cite{Ye:2022vbk, KM3Net:2016zxf}, insterstellar dust~\cite{POEMMA:2020ykm}, or mountains~\cite{Sasaki:2017zwd,Romero-Wolf:2020pzh,Brown:2021lef} as a neutrino target.
The subset of these that are deployed in liquid or solid water--hereafter referred to as water and ice respectively--have a long history~\cite{Spiering:2012xe}, and most of the features of current and future detectors can be traced back to the DUMAND project~\cite{RevModPhys.64.259}.
The largest currently operating of these detectors is the IceCube Neutrino Observatory~\cite{IceCube:2016zyt} located near the geographic South Pole.
Two other collaborations are currently constructing detectors: the KM3NeT collaboration is constructing ORCA and ARCA~\cite{KM3Net:2016zxf} in the Mediterranean Sea and the BDUNT collaboration is currently building Baikal-GVD in Lake Baikal in Russia~\cite{Avrorin:2015wba}. 
Additionally, two new experiments--P-ONE~\cite{P-ONE:2020ljt} off the coast of Vancouver in the Pacific Ocean and TRIDENT~\cite{Ye:2022vbk} in the South China Sea--and expansions of the IceCube Observatory~\cite{Ishihara:2019aao, IceCube-Gen2:2020qha} are under development.

Since all these experiments operate on the same detection principle and are deployed in water or ice, they share many technological features.
Each detector is comprised of individual optical modules (OMs) capable of detecting Cherenkov photons emitted by the charged byproducts of neutrino interactions. 
The arrangement and details of each OM vary from one detector to another depending on the optical properties of the medium, physics goals, and historical context of construction.

\begin{figure}[t]
  \centering
  \includegraphics[width=0.96\textwidth]{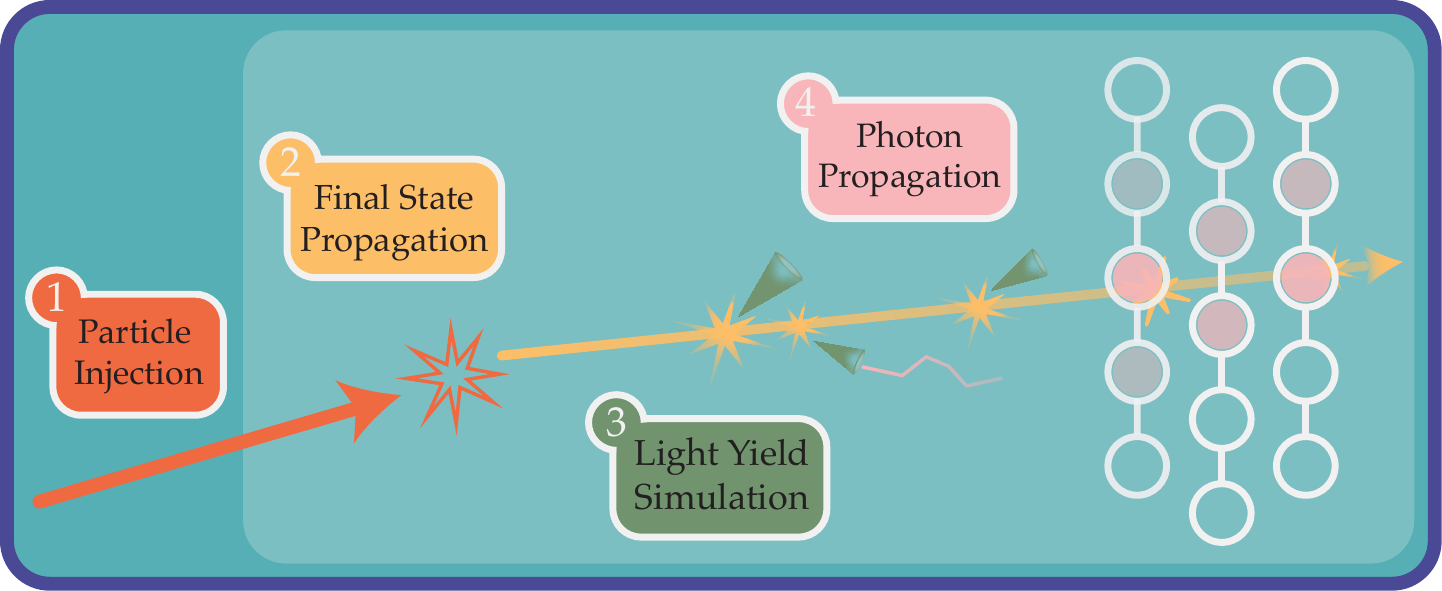}
  \caption{
    \textbf{\textit{Schematic showing the physical processes \prometheus{} models.}}
    (1), \prometheus{} selects an interaction vertex within \textit{simulation volume}, depicted here by the lighter-colored region.
    (2), the final states of this interaction are then propagated, accounting for energy losses and any daughter particles which may be produced.
    (3), these losses are then converted to a number of photons.
    (4), finally, these photons are then propagated until they either are absorbed or reach an optical module.
  }
  \label{fig:physics_process}
\end{figure}

Unsurprisingly, these commonalities result in similar simulation chains.
Most simulation chains follow some variation of the steps outlined in Fig.~\ref{fig:physics_process}, before eventually convolving the detector response with the distribution of photons that arrive at the OMs.
Since only this last step is not generic, there is an opportunity to develop a common software framework. 
\prometheus{} aspires to meet this opportunity by providing an integrated framework to simulate these common steps for arbitrary detectors in water and ice.
The flexibility allows one to optimize detector configurations for specific physics goals, while the common format allows one to develop reconstruction techniques that may be applied across different experiments.

The rest of this article is organized as follows.
In Sec.~\ref{sec:overview}, we provide a historical summary of the work that led to \prometheus{} and give an overview of \prometheus{}.
In Sec.~\ref{sec:examples}, we provide useful examples of code usage.
In Sec.~\ref{sec:output}, we describe the output format of \prometheus{}.
In Sec.~\ref{sec:performance}, we evaluate the performance of the code.
In Sec.~\ref{sec:checks}, we describe the benchmark checks performed on the code.
Finally, in Sec.~\ref{sec:conclusions} we conclude.

%% file: sections/overview.tex
\section{Overview of Software and of Relevant Physics}
\label{sec:overview}

\begin{figure*}[t]
  \centering
  \includegraphics[width=1\textwidth]{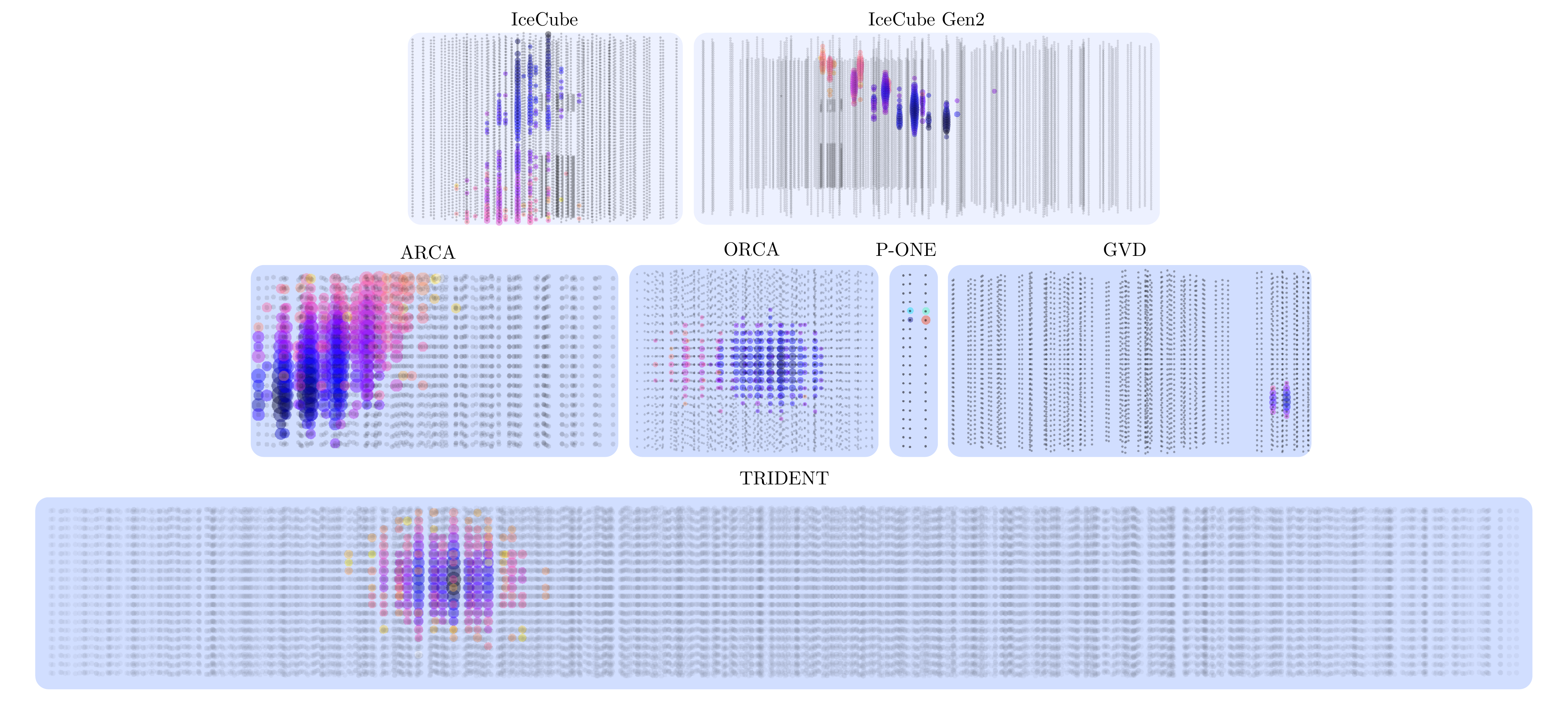}
  \caption{\textbf{\textit{Event views for various detector geometries.}}
  This shows the events created by either $\nu_{\mu}$ charged-current or $\nu_{e}$ charged-current interactions in a variety of geometries of current and proposed neutrino telescopes.
  Each black dot is an OM, while each colored dot indicates the average time at which photons arrived at the OM; black indicates an earlier arrival, orange indicates a later arrival, and purple an arrival in between.
  Furthermore, the size of the colored spheres is proportional to the number of photons that arrived at the OM.
  Detectors which appear against lighter blue backgrounds---the top row---are ice-based, while those against the darker blue backgrounds are water-based.
  }
\end{figure*}

\prometheus{} builds upon several decades of experience in the design of neutrino telescope simulations by using publicly available, well-maintained software whenever possible.
Neutrino event generation in these types of detectors can be traced to the first neutrino telescope event generator~\cite{osti_5884484,Hill:1996hzh,Gazizov:2004va,Yoshida:2003js}.
The first simulation of a neutrino telescopes in ice dates back to AMANDA and was called \texttt{NuSim}~\cite{Hill:1996hzh}, originally written in \texttt{Java}, while in water the earliest reference can be traced to ANTARES~\cite{Bailey:2002}.
\texttt{NuSim} was latter ported to \texttt{C++} and released as \texttt{ANIS}~\cite{Gazizov:2004va}.
This was then adapted into an internal IceCube event generator called \texttt{NuGen}~\cite{DeYoung:865626}.
Recently, the IceCube collaboration has released a new neutrino event generation that builds on these efforts called \texttt{LeptonInjector}~\cite{IceCube:2020tcq}; see Ref.~\cite{KM3NeT:2020tvi} for a similar effort in KM3NeT.
\texttt{LeptonInjector}~\cite{LeptonInjectorRepository} performs only neutrino injection around the detector and leaves neutrino transport through Earth as an \textit{a posteriori} weight~\cite{LeptonWeighterRepository} since it can be readily performed by packages such as those given in Refs.~\cite{Yoshida:2003js,Arguelles:2020hss,Vincent:2017svp,Safa:2021ghs,Garcia:2020jwr,Arguelles:2021twb,Garg:2022ugd}.

The propagation of high-energy muons is described in detail in Ref.~\cite{Lipari:1991ut}; see~\cite{Fedynitch:2021ima} for a recent revision.
Muon propagation in detailed Monte Carlo simulation was implemented in \texttt{MUSIC}~\cite{Antonioli:1997qw}, primarily used in water-based neutrino experiments, and \texttt{MMC}~\cite{Chirkin:2004hz} in ice-based experiments.
The latest and most up-to-date muon propagator optimized for neutrino telescopes is called \texttt{PROPOSAL}~\cite{Koehne:2013gpa} and builds on \texttt{MMC}.
\prometheus{} uses \texttt{PROPOSAL} to simulate the propagation of muons.
Tau propagation is also handled by \texttt{PROPOSAL}, though in most of the energy range of the experiments considered here, the tau losses are negligible; see~\cite{Bugaev:2002gy,Hagiwara:2003di,Dutta:2005yt,Bigas:2008ff,Arguelles:2022bma} for discussions on tau energy losses.

The emission of Cherenkov light from hadronic and electromagnetic showers in water is discussed extensively in Refs.~\cite{Wiebusch:thesis,Niess_2006}.
In these references, the emission of light was parameterized from dedicated GEANT4 simulations, which have been recently refined in~\cite{2012APh....38...53R,Radel:2012ij}.
The emission of light from hadronic or electromagnetic showers has been implemented in the \texttt{Cascade Monte Carlo} (\texttt{CMC}) package by the IceCube collaboration~\cite{IceCube:2020tcq} following the physics outlined in~\cite{Niess_2006}.
Unfortunately, \texttt{CMC} is not publicly available and is only usable in ice.
For this reason, we have re\"{i}mplemented the light yields produced by showers in \prometheus{} following the parameterizations given in~\cite{Niess_2006,2012APh....38...53R,Radel:2012ij}.
These are implemented in a module called \texttt{Fennel} in \prometheus{}, which is described in Sec.~\ref{subsec:light_yield}.

Finally, \prometheus{} solves the light transport problem using two different modules.
In the case of light propagation in ice, we use the standalone, open-source version of \texttt{PPC}, which is the ray tracer used by the IceCube collaboration and can be found in~\cite{Chirkin:2015kga,chirkin2022kpl}, while in the case of water, we implement our own ray tracing routines in a module called \texttt{Hyperion}, which is described in Sec.~\ref{subsec:photon_prop}.

These four steps---event injection, final state propagation, light yield simulation, and photon propagation--- each come with their own set of unique challenges.
As noted above, we will use publicly available and well-maintained packages to address these challenges whenever possible.
However, in \prometheus{} we take these challenges as opportunities to provide new solutions to them when publicly available software is lacking.
In what follows we will discuss the output of \prometheus{} and then outline the code structure and summarize the corresponding physics of each piece of code.

%% file: sections/examples.tex
\section{Examples}
\label{sec:examples}

In this section, we will lay out the basic use cases of simulating neutrino events in two detectors, one ice-based and one water-based.
This is meant to illustrate the basic configuration options for using a predefined detector.
This section should give the user a basic understanding and should enable them to simulate many physics scenarios of interest; however, this is by no means an exhaustive list of the configuration options, and we provide example code to handle many other scenarios in Appendix~\ref{app:examples}.

\subsection{$\nu_{\mu}$ Charged-Current Events in an Ice-Based Detector}

In this section, we will simulate $\nu_{\mu}$ charged-current interactions.
This particular example accounts for the ability of the resulting $\mu^{-}$ to travel large distances.
This will be discussed further in Sec.~\ref{subsec:lepton_injector}, and an example of disabling this option to simulate neutrinos whose interaction vertex is contained between the fiducial volume, so-called starting events, can be found in Ex.~\ref{ex:starting_events}

First, we will import \texttt{Prometheus} and define the resources and output directory relative to the \prometheus{} import.
\begin{lstlisting}[language=python,upquote=true]
import prometheus

prometheus_base = '/'.join(prometheus.__path__[0].split('/')[:-1])
resource_dir = f"{prometheus_base}/resources/"
output_dir = f"{prometheus_base}/examples/output/"
\end{lstlisting}
Next, we will import the \texttt{config} dictionary, the main interface for configuring \prometheus{}.
We set the run number, the number of events we want to simulate, and optionally, the random state seed.
\begin{lstlisting}[language=python, firstnumber=6]
from prometheus import config

config["run"]["run number"] = 925 # This will also be the random state seed unless the later line is uncommented
config["run"]["nevents"] = 100
config["run"]["storage prefix"] = output_dir
# Uncomment this line to set the state seed independently
# config["run"]["random state seed"] = 853
\end{lstlisting}
Now, we define the detector geometry.
For this example, we use a demonstration detector that we have made for this purpose.
This can be changed to simulate the approximate geometries of IceCube, the DeepCore subarray, the IceCube Upgrade, or IceCube Gen2 by changing \texttt{demo\_ice} to \texttt{icecube}, \texttt{deepcore}, \texttt{icecube\_upgrade}, \texttt{icecube\_gen2} respectively.
\begin{lstlisting}[language=python, firstnumber=13]
geofile = f"{resource_dir}/geofiles/demo_ice.geo"
config["detector"]["geo file"] = geofile
\end{lstlisting}
Next, we set the injection parameters.
Since we are simulating $\nu_{\mu}$ charged-current events, we will restrict ourselves to up-going events, where the Earth filters out atmospheric muons.
As we will discuss in Sec.~\ref{subsec:prelim}, we choose the direction vector to be aligned with the particle momentum.
Thus, events coming straight through the Earth have a zenith angle of $0^{\circ}$, and the horizon is located at $90^{\circ}$.
\begin{lstlisting}[language=python, firstnumber=13]
injector = "LeptonInjector"
config["injection"]["name"] = injector
injection_config = config["injection"][injector]
# Inject only upgoing events
degrees = 3.1415926536 / 180
injection_config["simulation"]["min zenith"] = 0 * degrees
injection_config["simulation"]["max zenith"] = 90 * degrees
# Inject with energies from 100 GeV to 1 Pev with energies sampled according to E^-1
injection_config["simulation"]["minimal energy"] = 1e2
injection_config["simulation"]["maximal energy"] = 1e6
injection_config["simulation"]["gamma"] = 1
# Consider only numu cc events
injection_config["simulation"]["final state 1"] = "MuMinus"
injection_config["simulation"]["final state 2"] = "Hadrons"
\end{lstlisting}
We can stipulate where the output should go, but it will also be handled internally if it is not specified.
The default behavior is to make a file called \texttt{{run number}\_photons.parquet}, in the output directory that we set earlier.
In case this is desired, we leave the commented-out code below as an example.
\begin{lstlisting}[language=python, firstnumber=29,upquote=true]
# Uncomment this to point to a preferred storage location.
# By default it goes to ....
# config["run"]["outfile"] = f"/path/to/your/folder/{config['run']['run number']}_photons.parquet"
\end{lstlisting}

Now we can simulate events!
\begin{lstlisting}[language=python,firstnumber=32,upquote=true]
from prometheus import Prometheus
p = Prometheus(config)
p.sim()
\end{lstlisting}

\subsection{$\bar\nu_{e}$ Neutral-Current Events in a Water-Based Detector}

Now we will inject neutral-current events from an incident $\bar\nu_{e}$ flux.
We will simulate all-sky neutrinos since one may use the outer layers of the detector to veto atmospheric $\mu^{\pm}$ when analyzing such events.
Much of this example will look familiar as it is similar to the prior example, and we will explain differences in the text whenever relevant.

Once again, we will first import \texttt{Prometheus} and point to the resources and output directory.
\begin{lstlisting}[language=python,upquote=true]
import prometheus
resource_dir = f"{'/'.join(prometheus.__path__[0].split('/')[:-1])}/resources/"
output_dir = f"{'/'.join(prometheus.__path__[0].split('/')[:-1])}/examples/output/"
\end{lstlisting}
Next, we will import the \texttt{config} dictionary and set the run number and the number of events we want to simulate.
Please note that the run number should be different each time since this is used when determining the output name if none is provided.
\begin{lstlisting}[language=python, firstnumber=4]
from prometheus import config

config["run"]["run number"] = 815 # This will also be the random state seed unless the later line is uncommented
config["run"]["nevents"] = 100
config["run"]["storage prefix"] = output_dir
# Uncomment this line to set the state seed independently of the run number
# config["run"]["random state seed"] = 853
\end{lstlisting}

Now, we set which detector we want to use, water-based this time.
\begin{lstlisting}[language=python, firstnumber=11]
geofile = f"{resource_dir}/geofiles/demo_water.geo"
config["detector"]["geo file"] = geofile
\end{lstlisting}
Next, we will set the injection parameters.
This time, we will ask for a neutral-current interaction coming from every direction in the sky.
Since these interactions are more time-consuming to simulate, we will sample according to a softer spectrum in comparison to the prior example.
\begin{lstlisting}[language=python, firstnumber=13]
injector = "LeptonInjector"
config["injection"]["name"] = injector
injection_config = config["injection"][injector]
# Inject only upgoing events
degrees = 3.1415926536 / 180
injection_config["simulation"]["min zenith"] = 0 * degrees
injection_config["simulation"]["max zenith"] = 180 * degrees
# Inject with energies from 100 GeV to 1 Pev with energies sampled according to E^-2
injection_config["simulation"]["minimal energy"] = 1e2
injection_config["simulation"]["maximal energy"] = 1e6
injection_config["simulation"]["gamma"] = 2
# Consider only nue nc events
injection_config["simulation"]["final state 1"] = "NuEBar"
injection_config["simulation"]["final state 2"] = "Hadrons"
\end{lstlisting}
Once again, we can stipulate where the output should go, but it will also be handled internally if it is not specified.
\begin{lstlisting}[language=python, firstnumber=27,upquote=true]
# Uncomment this to point to a preferred storage location.
# By default it goes to f"{config['run']['storage prefix']}/{config['run']['run number']}_photons.parquet"
# config["run"]["outfile"] = f"/path/to/your/folder/{config["run"]["run number"]}_photons.parquet"
\end{lstlisting}

Now we can simulate events!
\begin{lstlisting}[language=python, firstnumber=30,upquote=true]
from prometheus import Prometheus
p = Prometheus(config)
p.sim()
\end{lstlisting}

%% file: sections/output.tex
\section{Output}
\label{sec:output}

\prometheus{} events are serialized as \texttt{Parquet} files.
\texttt{Parquet} is a columnar format developed by Apache~\cite{parquet_docs} that supports nested data.
Columnar storage has been shown to yield improved performance when processing data and increased data storage compression~\cite{Blomer_2018}.
\texttt{Parquet} stores nested data structures using the technique introduced by Google in~\cite{36632}.

\begin{figure*}[tb]
  \centering
  \includegraphics[width=0.96\textwidth]{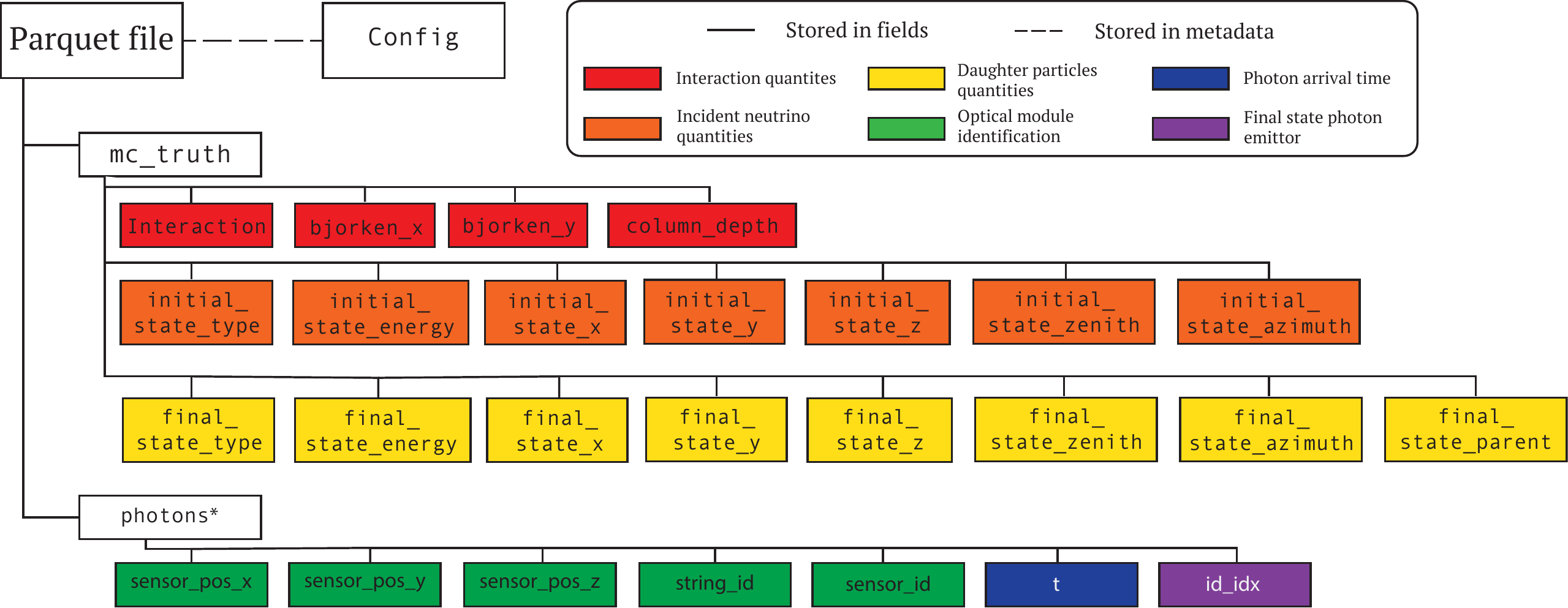}
  \caption{
  \textbf{\textit{Output format for default \texttt{Prometheus} \texttt{parquet} files.}}
  The solid lines indicate that information is stored in \texttt{field}s, while the dashed line indicates that information is stored in the metadata.
  We delay detailed discussion until Ex.~\ref{ex:output}, where we explain each field and compute basic quantities of interest.
  Fields with an asterisk can be renamed by the user to be compatible with legacy conventions.
  }
  \label{fig:codeflow}
\end{figure*}

When choosing the output format for \prometheus{} we have surveyed multiple options used in the community~\cite{Blomer_2018} such as \texttt{HDF5}~\cite{blomer2018quantitative} and \texttt{ROOT} ``n-tuples''~\cite{buckley2015implementation}.
Ref.~\cite{Blomer_2018}, studied different format disk usage and access speed in the context of collider experiment events.
They found that \texttt{Parquet} file size is comparable to \texttt{HDF5} and \texttt{ROOT} for uncompressed files and improved over the former when compressed by \texttt{zlib}.
They found that \texttt{Parquet} files read access per event is a factor of five times faster when compared to \texttt{HDF5} when using uncompressed \texttt{Parquet} files, while a factor of three when compared to the compressed version.
On the other hand, \texttt{Parquet} files have been shown to be a factor of three times slower than \texttt{ROOT} ``n-tuples'' when reading them.
We have opted not to use \texttt{ROOT} for interoperability reasons and to reduce dependence on additional libraries needed to work with \prometheus{} output.
Of the interoperable formats, we have decided to use \texttt{Parquet} over \texttt{HDF5} due to its improved performance as discussed above.

Here, we will broadly describe the information contained in the output files and the general structure, delaying a detailed example until Ex.~\ref{ex:output}.
The output files contain two fields, \texttt{mc\_truth} and \texttt{photons}.
For compatibility with naming conventions used in beta versions of the software, the first of these fields may be changed by the user; please see Appendix~\ref{app:config} for further details.

The \texttt{mc\_truth} field contains information about the injection quantities, such as the interaction vertex; interaction Bjorken variables; column depth traversed by the initial neutrino; the initial neutrino type, energy, and direction; and the final state types, energies, directions, and parent particles.
Since, in general, there can be any number of final-state particles, all final-state data are stored as one-dimensional arrays.
The order of the arrays is determined by traversing the MC tree of children, depth first.

The \texttt{photons} field contains information about photons that reach OMs.
This includes the OM identification numbers, OM position, photon arrival time, and an identification index that connects the photon to the final-state particle that created it.
If available, the photon arrival direction and position of the photon on the OM will also be saved.
This availability depends on which photon propagator is being used; please see Sec.~\ref{subsec:photon_prop} for further details.
These last two data can be useful for, \textit{e.g.}, simulating the OM acceptance in cases where it is heavily directionally dependent.

The configuration information is also stored in the \texttt{Parquet} file as metadata.
Once extracted, this may be dumped to a \texttt{json} file, and fed back into \prometheus{} to resimulate with the same parameters; please see Ex.~\ref{ex:output} for an example of this process.
This may be useful if you want to simulate the same event in different detectors in order to compare performance.
We should remark that while most of the code can be seeded to ensure reproducibility, \texttt{PPC} does not allow for seeding.
Previously there was a compile-time option to set the random state of \texttt{PPC}, but currently the random state is set using the time of day~\cite{dima_private_1}.
Thus, while simulations of water-based detectors are exactly reproducible, simulations for ice-based detectors will produce results that vary within Poisson fluctuations.

In addition to the main \texttt{Parquet} output, some steps in the \prometheus{} chain produce intermediate files.
We delete these files unless they contain information not available in the final output.
Currently, only the \leptoninjector{} \texttt{.lic} files, which contain the configuration information used in injection, meet this criterion.
These files are needed in order to weight events to obtain an event rate.
While it is possible to regenerate these after the fact, please do not remove these files if you intend to weight events.

%% file: sections/structure.tex
\section{Code Structure}
\label{sec:structure}

\subsection{Preliminary Remarks and Conventions}
\label{subsec:prelim}

\begin{figure*}[tb]
  \centering
  \includegraphics[width=0.96\textwidth]{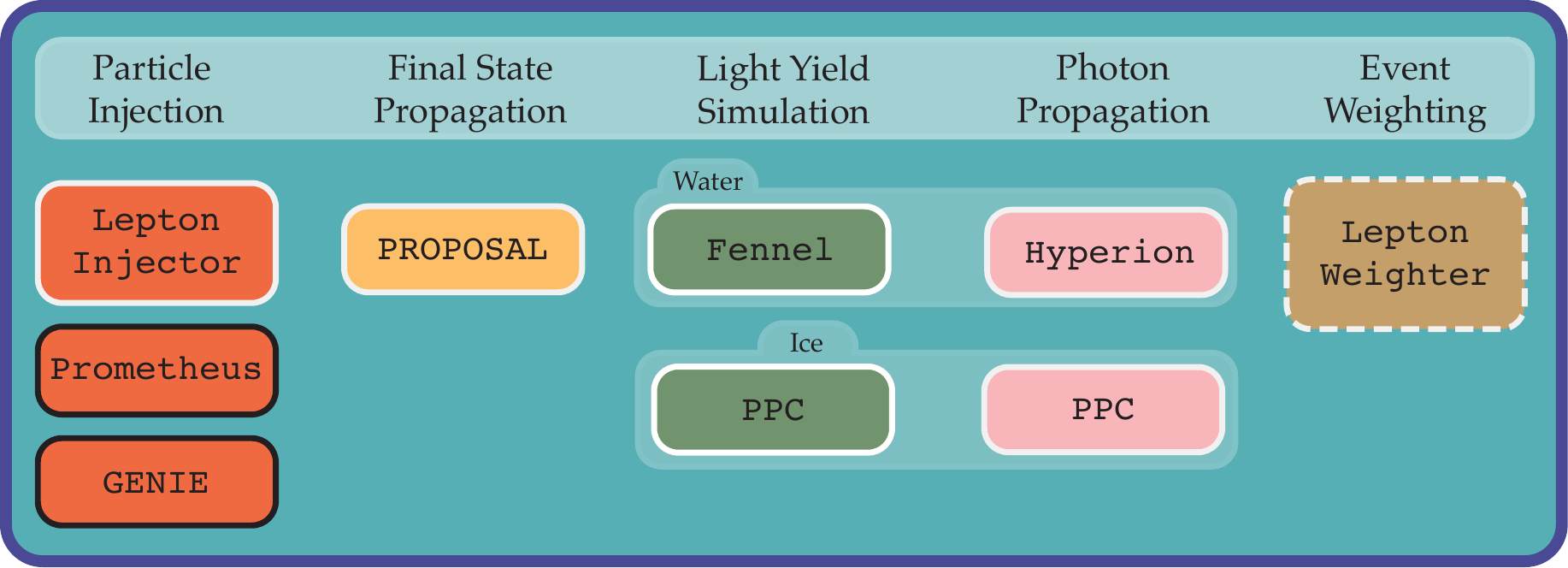}
  \caption{
  \textbf{\textit{Summary of packages used for different stages in the code.}}
  The boxes outlined in white are the default packages used, while boxes outlined in black have optional interfaces.
  Event weighting has a dashed outline to denote that this step is optional.
  The default behavior of the light yield calculation and the photon propagation depends on the medium, as is shown by the light shaded regions.
  }
  \label{fig:codeflow}
\end{figure*}

At its core, \prometheus{} is a framework for shepherding particles through the steps outlined in Fig.~\ref{fig:codeflow} in a consistent manner.
In the following sections, we will outline the \prometheus{} dataclasses that allow for this consistent treatment and explain the interfaces between \prometheus{} and the external packages.
While this is not comprehensive, it should give a sufficient understanding to work with the package.
We will point the reader to references that describe the external packages in more detail when appropriate.
Along the way, we will point out ways to adapt \prometheus{} to different simulations needs, including extending \prometheus{} to work with additional external packages and discussing user-configurable parameters.
We will refrain from discussing the ``how" of using these parameters, leaving that to examples in Sec.~\ref{sec:examples}, preferring instead to describe the impact of the parameters on the simulation.
Unless it is noted otherwise, all parameters can be found in the \texttt{config} file.
While we will not be able to describe every available, configurable parameter, we have a comprehensive list---including descriptions, location in the \texttt{config} file, and defaults---in Appendix~\ref{app:config}.

Before embarking, we should first discuss \prometheus{} conventions.
\prometheus{} uses a unit system where the units of length, energy, time, and angle are the meter, GeV, nanosecond, and radian respectively.
This means that all user input should be provided in these units and that all output will be provided in these units.
When interfacing new, external packages, one should account for unit conversions from \prometheus{} units to the units of the new package.
Furthermore, we follow a convention where the direction vector is aligned with the momentum of the particle.
Thus, up-going events result from neutrinos with incident direction between $0^{\circ}$ and $90^{\circ}$ and down-going events have directions between $90^{\circ}$ and $180^{\circ}$.
This is the opposite convention that many observatories use, where the direction is anti-aligned with the particle momentum, thus describing where the neutrino originated.
To better interface with the larger high-energy physics community, we have chosen the convention that is broadly used in accelerator neutrino experiments.

\subsection{\particle{} Dataclass}

The fundamental dataclass of \prometheus{} is the \particle{}, which minimally contains the particle type, an integer following the Particle Data Group (PDG) convention given in~\cite{Zyla:2020zbs}; the energy of the particle at creation; the direction in which the particle is travelling; and the position of the particle that is relevant to the simulations.
This position may be either the interaction vertex, as is the case for incident neutrinos, or the point of creation, as is the case for secondary particles.
While it is the case that for many situations of interest, the creation point of the secondary particles will overlap with the interaction vertex, this is not always the case.
For instance, in the case of $\nu_{\tau}$ charged-current interactions, the final state $\tau^{\pm}$ may decay at a point significantly offset from the interaction vertex, producing charged particles and leading to unique event signatures~\cite{IceCube:2020fpi}.
Furthermore, some injection packages give detailed particle output from the initial interaction, and some of these particles may be created offset from the interaction vertex.
Thus, we need to make this distinction in the definition of the position in order to accommodate these situations.

We resolve this ambiguity by defining a subclass of the \particle{}, the \propagatableparticle{}, which tracks all particles that can generate energy losses and ultimately light.
It is precisely these particles that may be created at an offset from the interaction vertex.
Thus, the position of any instance of this class is the point at which it was created, while the position of a \particle{} that is not an instance of this class, is the interaction vertex.
In addition to this distinction, this subclass has four new attributes, \texttt{losses}, which tracks energy losses of the particle; \texttt{parent} which is a \particle{} or \propagatableparticle{} object; \texttt{children}, a potentially empty list of \propagatableparticle{} objects; and \texttt{hits}, a list of all photons produced by the \propagatableparticle{} that hit an optical module before being absorbed.

\subsection{\detector{} Configuration}
\label{subsec:det}

The information about the position of the OMs and the propagation medium is contained in the \texttt{Detector} object.
The coordinate system in which the OM positions are specified has its origin at the water-air interface.
Only the relative $x$- and $y$-coordinates will affect the simulation, but the $z$-coordinate plays a crucial role in particle injection and charged lepton propagation.
The choice of medium specifies which light yield calculation and photon propagation calculations will be used.
In the case of ice, \texttt{PPC} will be used to calculate both quantities, while for water, \texttt{fennel} and \texttt{Hyperion} will be used to calculate the light yield and photon propagation respectively.
See Secs.~\ref{subsec:light_yield} and~\ref{subsec:photon_prop} for more details on these calculations.

While the user may manually specify the position and other properties of each OM, we provide several potentially more expedient ways to specify detectors.
The first is through \prometheus{} \texttt{geo} files.
These are text files with a specified format that, at minimum, give the locations of all optical modules and the detector medium.
Although this is the only required information, they may include any additional detector metadata as needed.

\prometheus{} provides \texttt{geo} files with approximate OM locations for the IceCube, IceCube Upgrade, IceCube Gen-2, ORCA, ARCA, Baikal-GVD, P-ONE, and TRIDENT detectors, and each has an associated Earth model that will be used for particle injection and charged lepton propagation.
See Appendix~\ref{app:earth_models} for a more detailed discussion of the Earth models used for each detector.

In addition to specifying the detector via provided \texttt{geo} files, \prometheus{} provides utilities for generating new detector geometries.
These include utilities to make a line of vertically aligned OMs, or a triangularly, hexagonally, orthogonally, or rhombically arranged set of such lines.
These lines may then be combined using \texttt{Python}'s built-in \texttt{+} function.
This may be convenient for, \textit{e.g.}, designing detectors in the vein of Baikal-GVD or P-ONE which are made up of a number of identical clusters.
Detectors constructed in this way can then be exported to a \texttt{geo} for later use.
See Ex.~\ref{ex:detector_build} for an example of building and a detector in this manner.
While custom Earth models may be made for such detectors, we include two generic Earth models.
For ice-based detectors, we use a generic South Pole Earth model from~\cite{LeptonInjectorRepository}, and for water-based detectors, we use the PREM model with 2~km of water appended.

\subsection{Primary Particle \texttt{Injection}}

As discussed above, injection is the process of forcing a neutrino to interact, creating particles that may produce light, and thus trigger the detector.
This requires balancing the need to simulate all interactions that may cause the detector to trigger while not wasting computational resources simulating events which have a negligible chance of doing so.
While this problem has been addressed by a number packages, such as~\cite{Andreopoulos:2015wxa,IceCube:2020tcq,Garcia:2020jwr}.
We will spend limited words describing the approaches these packages take to solve this problem focusing instead on injection options available in \prometheus{} and possibilities for extending these options.

\subsubsection{Default Injection: \texttt{LeptonInjector}}
\label{subsec:lepton_injector}

By default, \prometheus{} uses \leptoninjector{}~\cite{IceCube:2020tcq} to select the interaction vertex, initial neutrino energy and direction, and final-state energies.
The energy sampling is done according to a power-law with an index of $\gamma$ in the range $E \in \left[E_{\rm{min}},\,E_{\rm{max}}\right]$.
The incident direction is sampled uniformly in phase space, \textit{i.e.}, uniformly in the azimuthal angle and uniformly in the cosine of the zenith angle.
The azimuthal angle will lie in $\phi\in\left[\phi_{\rm{min}}, \phi_{\rm{max}}\right]$ and the zenith lies with $\theta \in \left[\theta_{\rm{min}},\,\theta_{\rm{max}}\right]$.
The parameters $\gamma$, $E_{\rm{min}}$, $E_{\rm{max}}$, $\phi_{\rm{min}}$, $\phi_{\rm{max}}$, $\theta_{\rm{min}}$, and $\theta_{\rm{max}}$ can be set by the user.
See Appendix~\ref{app:config} for more details on this.

The interaction vertex can be sampled in one of two ways: \texttt{RangedInjection} or \texttt{VolumeInjection}.
These terms are described in detail in~\cite{IceCube:2020tcq}, but we will briefly summarize them and introduce relevant variables here.
At energies above 1~TeV, $\mu^{\pm}$, $\tau^{\pm}$, and some of their daughter leptons, can travel distances $\gtrsim 1$~km before stopping.
This distance grows as the energy of the charged lepton increases, and as such the effective volume of the detector grows with increasing energy.
\texttt{RangedInjection} accounts for this phenomenon and samples the distance between the interaction vertex and the detector in a manner appropriate to the particle energy.
The maximum radius of closest approach, $r_{\rm{inj}}$, and padding beyond the particle range, $\ell_{\rm{ec}}$ additionally affect the injection region.
\texttt{VolumeInjection} on the other hand, selects the interaction vertex within a cylinder with a symmetry axis aligned with the detector center of gravity in the $xy$-plane, and with radius and height $r_{\rm{cyl}}$ and $h_{\rm{cyl}}$.
This may be useful for simulating $\nu_{e}$ charged-current events, $\nu_{\alpha}$ neutral-current events, or $\nu_{\alpha}$ starting events.
The parameters $r_{\rm{inj}}$, $\ell_{\rm{ec}}$, $r_{\rm{cyl}}$, and $h_{\rm{cyl}}$ may be set by the user, but we consider these advanced injection options, and by default we will select values that will sample the full injection space, accounting for the scattering and absorption of the medium.

Currently, \leptoninjector{} cross-section tables only support energies down to 100~GeV.
If users wish to simulate neutrino events below this energy, please refer to the next section which describes providing injection from another software and using \prometheus{} to propagate the final-state particles.

\subsubsection{\texttt{Injection} Extensibility}

While new injection can be generated only by \leptoninjector{}, \prometheus{} can accept injection files from other sources.
The current iteration of the code provides interfaces for using both \prometheus{} and \texttt{GENIE} output files, and any files which take the form of \leptoninjector{} output.
The first may be useful for simulating the same events in different detectors in order to compare detector performance, while the second is useful for energies lower than \leptoninjector{}'s 100~GeV threshold.
The last may be used to simulate exotic physics scenarios not supported by standard neutrino injectors.
In order to access this feature, one needs only change the injector name, and then supply the name of the injection file in the appropriate field of the configuration file.
See Appendix~\ref{app:config} for further details.

We expect that these three options should satisfy almost all needs; however, if a new use case arises that cannot be accommodated, adding a new injection interface is fairly straightforward if tedious to describe in words and varies significantly depending on the injection file format.
We will refrain from doing so here, but welcome any user that should find themselves in this situation to contact the authors to avail themselves of this feature.

\subsection{Secondary Particle Propagation}
\label{subsec:secondary}

Once the final states have been generated in the injection step, the resulting particles must be propagated, accounting for energy losses and particle decay.
We assume that $K^{0}$, $K^{\pm}$, $\pi^{0}$, and $\pi^{\pm}$ begin depositing energy immediately, and as such do not propagate them beyond the point of creation.
Furthermore, we assume that all final-state neutrinos do not interact again near the detector, and may safely be ignored.
This is a safe assumption since the neutrino interaction length is $\sim10^7$~mwe at $10^6$~GeV.

In order to propagate final-state charged leptons, we rely on the \texttt{PROPOSAL} package~\cite{koehne2013proposal}.
This Monte-Carlo-based propagation package includes up-to-date cross sections for ionization; bremsstrahlung; photonuclear interactions; electron pair production; the Landau–Pomeranchuk–Migdal and Ter-Mikaelian effects; muon and tau decay; and Moli\`{e}re scattering.
While developing \prometheus{}, the most recent versions of \texttt{PROPOSAL} occasionally had trouble being installed via \texttt{pip} on certain operating systems.
To accommodate these issues, \prometheus{} has interfaces to run with \texttt{PROPOSAL v6.1.6} or \texttt{PROPOSAL v7.x.x}; however, at the time of writing, these installations issues have been resolved, and so we strongly recommend running with using the latest version.

We have decided not to expose all of \texttt{PROPOSAL}'s options to the  \prometheus{} user, preferring to restrict our options to those which have the largest impact in order to simplify the configuration experience.
If a use case requires an option that is not available by default, they may define a function to create a \texttt{PROPOSAL} propagator according to their needs.
This may then be interfaced with the appropriate \texttt{LeptonPropagator} class in \prometheus{} to supply the desired results.
Furthermore, if a user desires to use a different package to propagate leptons, this may be accomplished by creating a new \texttt{LeptonPropagator} class and implementing the appropriate abstract methods.

The medium in which the propagation takes place is governed by the same Earth model from the injection step.
For the final state propagation, however, we convert the layers given by the PREM, which are fit to a degree-three polynomial, to a layer of constant denisty.
The value of the density is the mean of the density at either end of the layer.
While this approximation breaks down near the center of the Earth, it holds in the region within 100~km of the detector.
This is roughly the maximum range of a charged particle, and so it is only in this region where the approximation holds that final state propagation should be taking place.
Please see Appendix~\ref{app:earth_models} for further discussion of this.

\begin{figure}[b]
  \centering
  \includegraphics[width=0.5\textwidth]{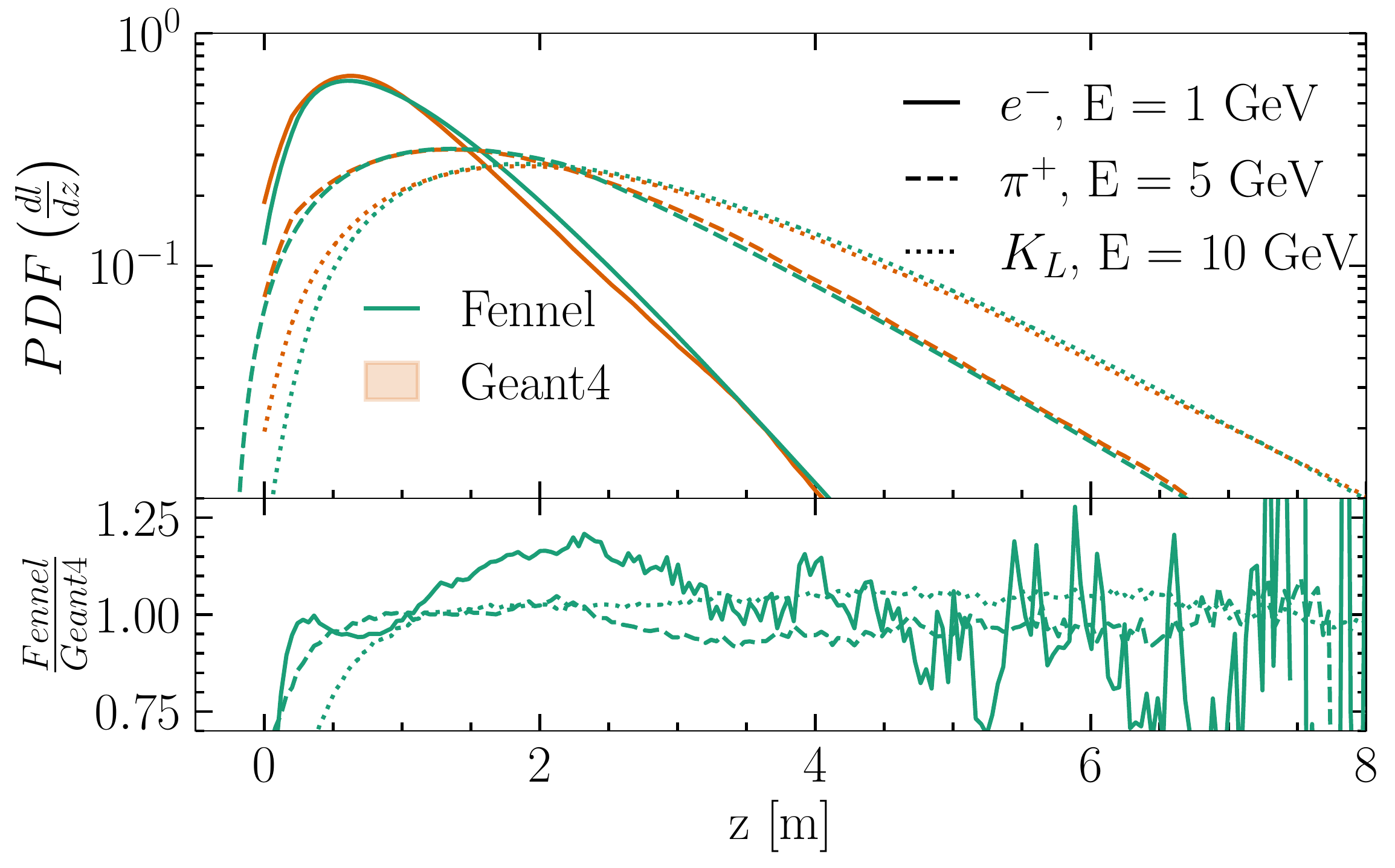}
  \caption{\textbf{\textit{Comparison between \texttt{GEANT4} and \texttt{fennel} longitudinal profiles.}} Top: Shown are the differential track lengths for three particle showers each with a different energies.
  Note, the shift of the maximum depending on energy. Bottom: The ratio of the differential track lengths.
  For most of the shower's development, the differences between \texttt{GEANT4} and \texttt{fennel} are below 20\%.
  }
  \label{fig:FennelvsGeant}
\end{figure}

\subsection{Light Yield Simulation}
\label{subsec:light_yield}

After propagating the final states, \prometheus{} must convert the energy deposited in and around the detector to photons.
\prometheus{} uses two different packages, depending on whether the detector being simulated is in water or ice.
In the former case, we use \texttt{fennel}, a new package developed for this work, while in the latter case, we use a standalone version of \texttt{PPC}~\cite{chirkin2022kpl}.
These both employ parameterizations of dedicated \texttt{GEANT4} simulations across a variety of energies.

\subsubsection{\texttt{fennel} For Water-Based Detectors}
When modeling neutrino detectors in water, \texttt{fennel}~\cite{fennel2022@github} is used to calculate the light yields and emission angles for the different losses occurring along a track, and from hadronic showers.
It utilizes the parametrizations described in \cite{Radel:2012ij} to quickly model the Cherenkov light produced by particles and their secondaries.
The parametrizations were produced by fitting \texttt{GEANT4}~\cite{GEANT4:2002zbu} shower distributions.
The relevant distribution for light yields is the total track length of charged particles in the triggered shower above the Cherenkov threshold.
A comparison between \texttt{fennel} and \texttt{GEANT4} is shown in Fig.~\ref{fig:FennelvsGeant}.
There we are comparing simulated electron, $\pi^+$, and $K_L$ showers with 1, 5, and 10~GeV energies, respectively.
Shown is the differential track length, $l$, depending on the shower depth $z$: $\frac{\mathrm{d}l}{\mathrm{d}z}$.
For the most significant regions for light emission, the difference between the parametrization and MC simulation is less than 20\%.
The photons are then handed over to \texttt{hyperion} for propagation.

\subsubsection{\texttt{Photon Propagation Code} For Ice-Based Detectors}
\label{subsubsec:ppc_light_yield}

When modeling neutrino detectors in ice, \texttt{PPC}~\cite{chirkin:2011abc,chirkin2022kpl} is used to calculate the number and angular distribution of photons from electromagnetic (EM) and hadronic cascades.
Internally, \texttt{PPC} bases the EM photon yield on the parameterization from~\cite{Radel:2012ij}, the same as \texttt{fennel}.
In this work, the simulated the photon yield and angular distribution from $e^{-}$, $e^{+}$, and $\gamma$---with energies ranging from 1~GeV to 10~TeV---in both ice and water using \texttt{GEANT4}.
The resulting longitudinal distributions were then fit to a known functional form.
The parameters of this fit for each EM particle agree within a factor of $10^{-3}$, \textit{i.e.}, the light yield for all EM particles is the same up to one part in one-thousand.
The light yield for hadronic showers is calculated by rescaling the EM photon yield per unit length by a constant which varies for each type hadron.

Within \texttt{PPC}, the distance from the start of the cascade to the point of photon emission is sampled from a $\Gamma$ distribution in order to properly account for the longitudinal development of the cascade.
Details on the parameterization and fit values used in \texttt{PPC} can be found in~\cite{Radel:2012ij}.
The photons resulting from this are then propagated internally by \texttt{PPC}.
This is described in Sec.~\ref{subsubsec:ppc_prop}.

\subsection{Photon Propagation}
\label{subsec:photon_prop}

The photons generated in the light yield calculation must finally be propagated.
This is usually solved via ray tracing of the photon until it is either absorbed or reaches an OM; however, if the Green's function of a photon reaching an OM is known, this may also be used to compute the transmission probability.
One may then use the accept-reject method to determine if the OM ``sees'' the photon.

As is the case for the light-yield calculation, \prometheus{} uses a different package depending on whether the detector is in water or ice.
In the former case, \prometheus{} uses \texttt{hyperion}, developed for this package, and can take advantage of the Green's function approach.
In the latter case, \prometheus{} uses the same open-source version of \texttt{PPC} which is used to compute the photon yield and which only uses the ray-tracing approach.

\subsubsection{\texttt{hyperion}  for water}
\label{subsec:hyperion}


\texttt{Hyperion} is used to propagate photons in water and without additional input, uses a Monte Carlo approach to do so.
Photons are represented as photon states, which include information about the photon's current position, its direction, time (or distance) since emission, and wavelength. Photons are initialized by drawing the initial state from distributions that represent the photon emission spectrum for the source class to be simulated.
The propagation loop involves three main steps: 1) Sample the distance to the next scattering step from an exponential distribution.
2) Calculate whether the photon path (given by the current photon position, distance to the next scattering step, and photon direction) intersects with a detector module.
3a) In case of intersection - the photon is stopped and its intersection position is recorded in the photon state.
3b) In case of no intersection - the photon is propagated to the next scattering site and a new direction is sampled using the scattering angle distribution.





\subsubsection{\texttt{Photon Propagation Code}}
\label{subsubsec:ppc_prop}

In addition to handling the photon yield in ice, \texttt{PPC} also carries out the photon propagation.
\texttt{PPC} uses Monte Carlo methods to propagate the photons until either they reach an optical module or they are absorbed.

The settings by which \texttt{PPC} propagates the photons may be set by a number of tables contained in specially-named text files.
These tables set the angular acceptance, size, and efficiency of the OMs; the mean deflection angle of a scattered photon; the depth and wavelength dependence of the scattering and absorption; and the so-called ``tilt'' of the ice.
In the \texttt{/resources/PPC\_tables/south\_pole/}, we provide tables that parametrize the ice beneath the south pole.
This uses scattering and absorption taken from~\cite{Rongen:2019wsh}, a uniform angular acceptance, and no tilt paramterization, \textit{i.e.} it assumes flat ice.
More details than these are known about the South Pole ice, \textit{e.g.} birefringence~\cite{Chirkin:2019vyq} and a non-zero-tilt~\cite{Rongen:2019wsh}; however, these details are very difficult to reconstruct without access to internal information.
Furthermore, the parametrization should provide and sufficiently accurate representation of the ice.

\subsection{Event \texttt{Weighting}}

While the simulated events can be generated according to arbitrary user input, these can be reweighted to a physical flux.
\prometheus{} does this via the \texttt{LeptonWeighter}~\cite{IceCube:2020tcq} package by computing the \texttt{oneweight}.
This quantity removes all the generation choices that were made when producing events, and, when multiplied by a desired flux, gives the rate for that neutrino event.
Thus, this may be used to reweight to any flux.
If the use case does not require physical rates, as is the case in many machine learning applications, weighting is not necessary.
As such, we do not perform weighting by default.
This may be done after the rest of the simulation has run with either the \leptoninjector{} \texttt{HDF5} files or the \prometheus{} \texttt{parquet} files through the \texttt{H5Weighter} and \texttt{ParquetWeighter} objects respectively.

To weight events, we use the \texttt{LeptonWeighter} package.
This is the companion to the \leptoninjector{} package, and requires the \texttt{lic} configuration files that are output at the \texttt{injection} step.
In addition to the \texttt{lic} file, \texttt{LeptonWeighter} needs to be provided differential cross-section files.

These files may be specified by the user; however, if the cross-section files are not provided, \prometheus{} will attempt to find suitable files.
The procedure for this depends on whether one is weighting from a \texttt{parquet} file or a \texttt{h5} file.
In the former case, it will search for appropriate cross section files in the directory provided in at configuration time, since this information is stored in the \texttt{parquet} file.
If this fails, then it will default to using the cross sections provided in the \prometheus{} resources directory.
In the latter case, \prometheus{} will use the cross sections provided in the resources directory directly.


%% file: sections/performance.tex
\section{Performance}
\label{sec:performance}

\subsection{Timing}
\label{subsec:timing}

\begin{figure}[bt]
  \centering
  \includegraphics[width=0.95\textwidth]{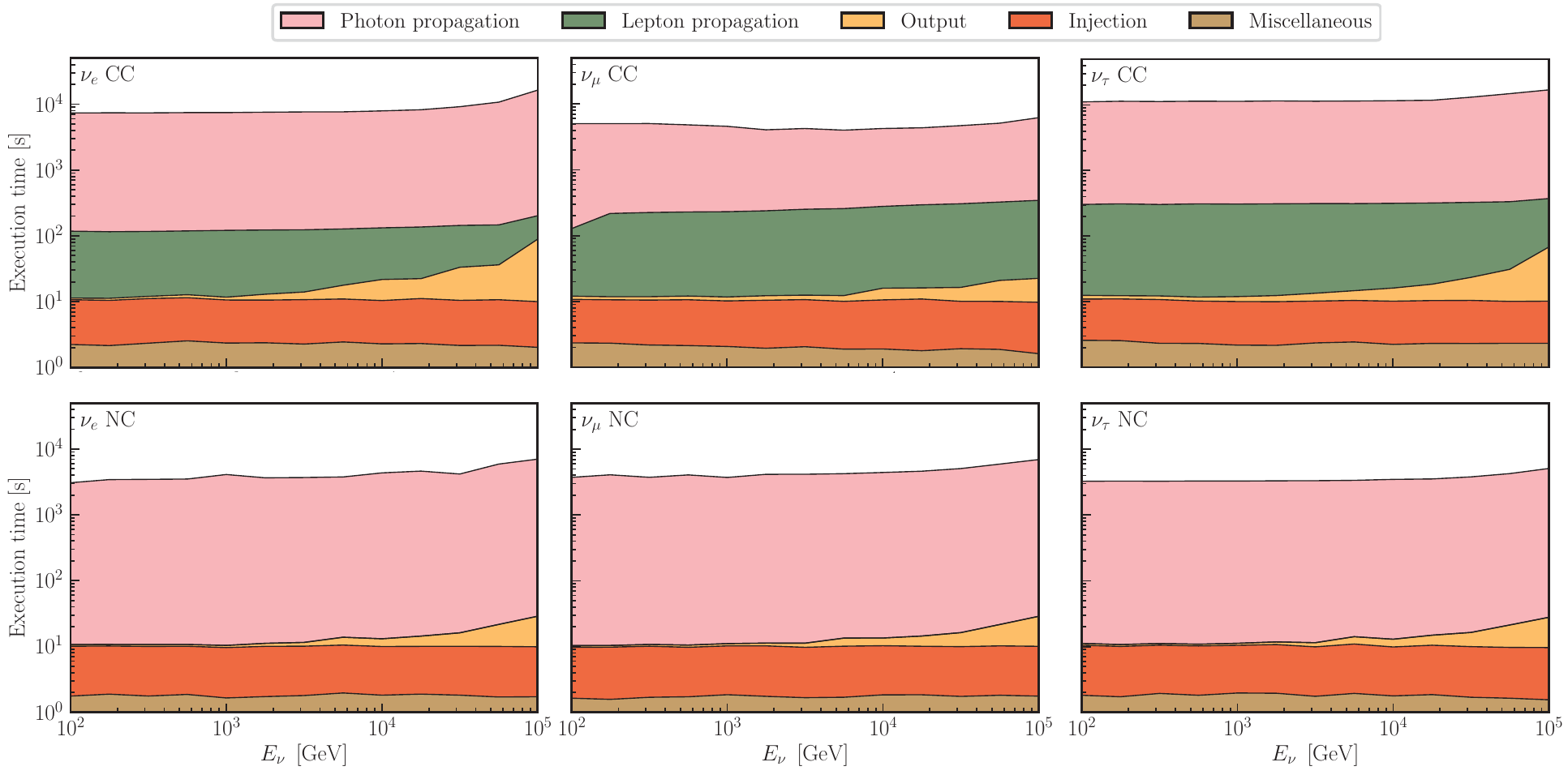}
  \caption{
    \textbf{\textit{Wall time per 1000 events as a function of neutrino energy.}}
    The runtime for thousand events propagated in ice using \texttt{PPC}.
    As expected, higher incident neutrino energies require longer run time.
    The colors, each corresponding to a different stage of production, are stacked.
    Each panel shows the runtime for different a different interaction type.
  }
  \label{fig:ice_timing}
\end{figure}

\begin{figure}[tb]
  \centering
  \includegraphics[width=0.95\textwidth]{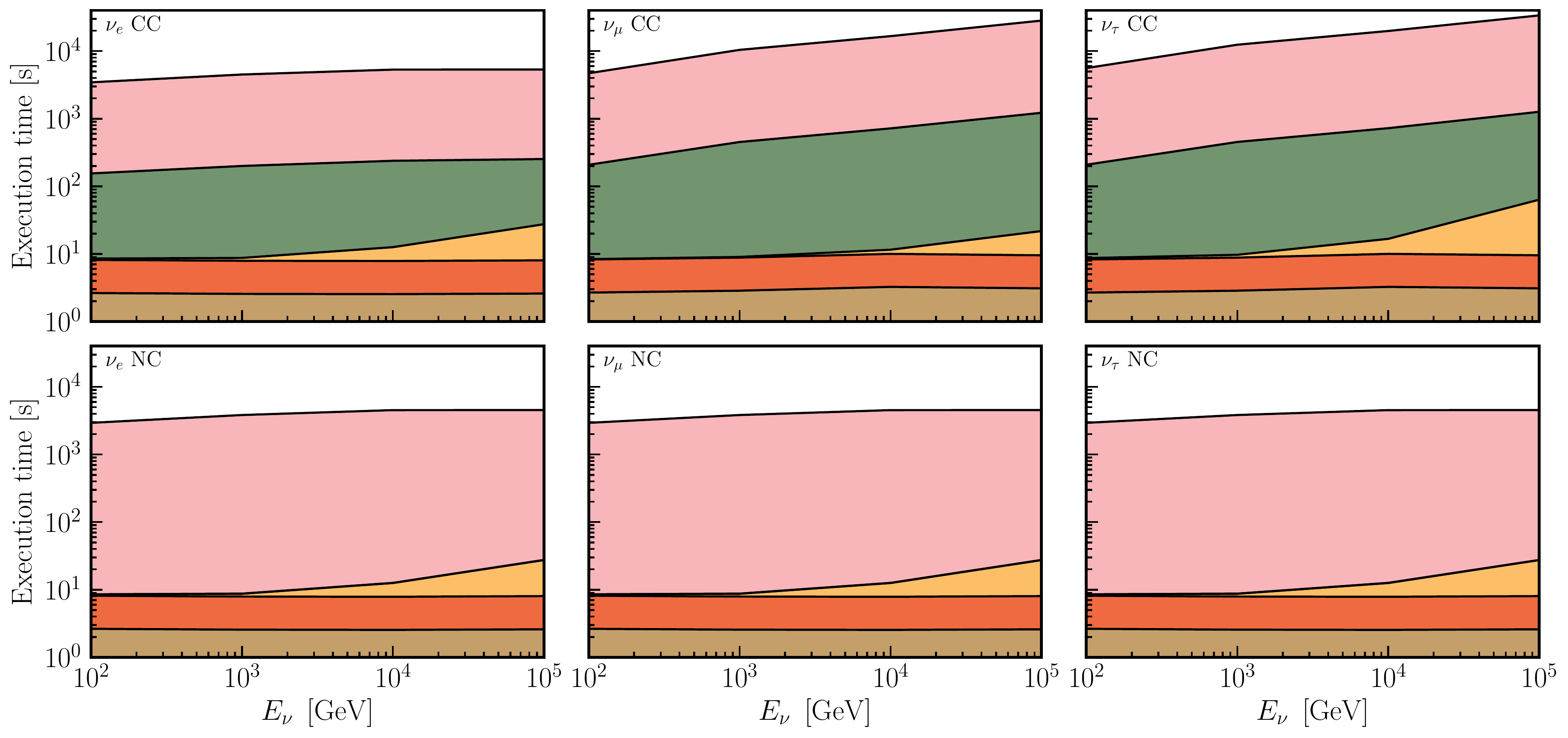}
  \caption{
    \textbf{\textit{Wall time per 1000 events as a function of neutrino energy.}}
    The runtime for thousand events propagated in water using \texttt{fennel} and \texttt{Hyperion}.
    As expected, higher incident neutrino energies require longer run time.
    The colors, each corresponding to a different stage of production, are stacked.
    Each panel shows the runtime for different a different interaction type.
  }
  \label{fig:water_timing}
\end{figure}

In order to quantify the timing performance of \prometheus{}, we ran the full simulation chain on example ice-based and water-based detectors, introduced in~\ref{sec:examples}.
For each detector, we simulated $10^{3}$ events for each flavor and interaction type combination at 13 energies equally log-spaced between $10^{2}$~GeV and $10^{5}$~GeV.
For the ice-based detectors, this test was run on a partition of an NVIDIA A100 GPU.
This partition has 10~GB of CPU memory and 10~GB of GPU memory.
The results of these tests for the ice-based detector can be seen in Fig.~\ref{fig:ice_timing}, while the results for the water-based detector can be found in Fig.~\ref{fig:water_timing}. The water test was run on a 12th Gen Intel(R) Core(TM) i7-1255U, with 16 GB CPU memory.

Some interesting trends that shed light both on the code and the underlying physics can be observed in these plots.
As one can see, photon propagation is the most time-consuming part of the simulation chain, taking up $\gtrsim$95\% of the time.
The next leading contribution is the charged lepton propagation.
As one would expect, this only contributes to the runtime when charged-current interactions are simulated since charged leptons are not produced in neutral-current interactions.
This is because, as discussed in Sec.~\ref{subsec:secondary}, we assume the neutrino emerging from a neutral-current interaction will not interact within the instrumented region, and thus, they do not require additional computational resources.
Careful observation reveals that the time required for this changes depending on whether $\nu_{e}$, $\nu_{\mu}$, or $\nu_{\tau}$ are simulated.
This is because in the first case a propagator only needs to be made for $e^{-}$, while for the other cases, a propagator must be made for at least two charged leptons since the primary product can decay to a lighter charged lepton.
Furthermore, it is worth noting that the initial particle injection and miscellaneous overhead do not scale with energy.
This is sensible since injection primarily relies on random number generation, and there is a limited amount of computational difference across energies.

In the case of a water-based detector, we see a similar pattern where the photon propagator is the main driver of the runtime.
Furthermore, we once again see the subdominant contribution of charged lepton propagation only in charged-current interactions.
Once again, the overhead time is much larger for $\nu_{\mu}$ and $\nu_{\tau}$ due to the need to create multiple propagators.

In summary, the runtime is dominated by the photon propagation, implying the importance of researching methods for accelerating this process, such as those that have been proposed in~\cite{Collin:2018oba}.

\subsection{Output Memory Usage}
\label{subsec:output_memory}

\begin{figure}[bt]
  \centering
  \includegraphics[width=0.7\textwidth]{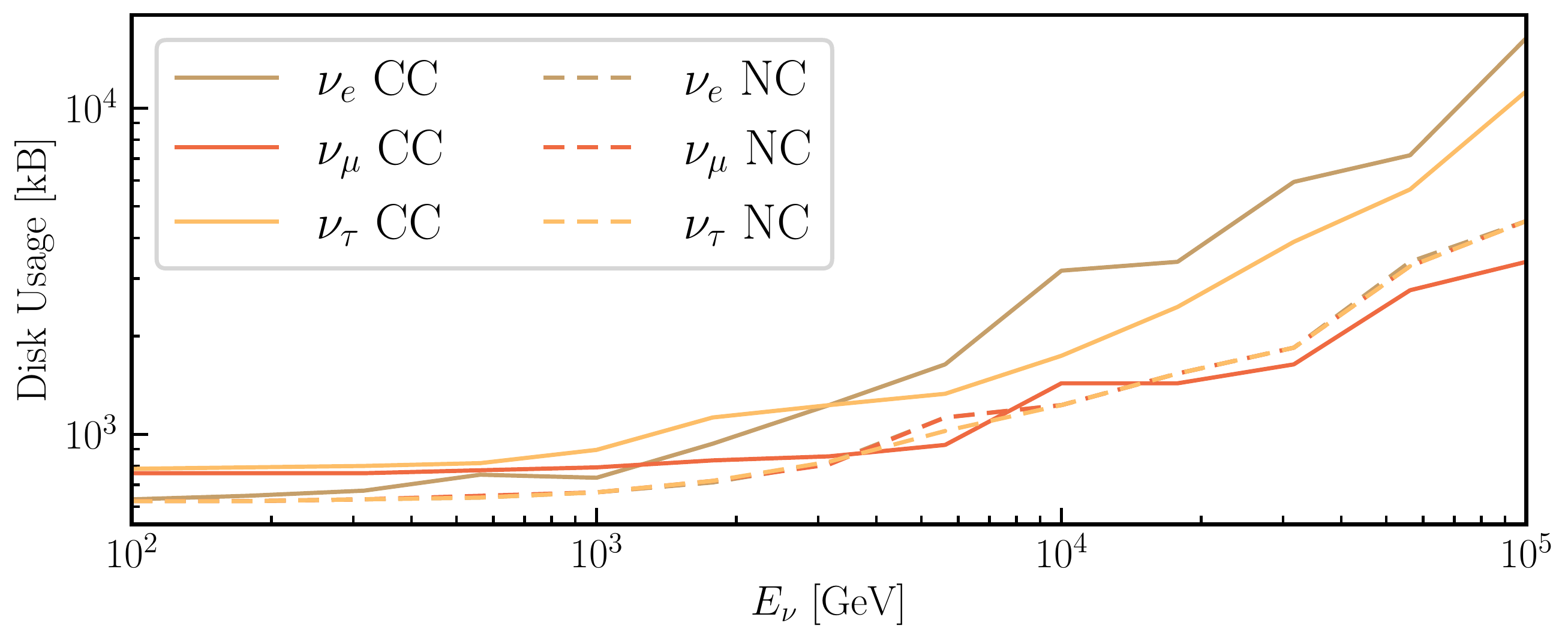}
  \caption{\textbf{\textit{Disk space per thousand events as a function of energy.}}
  The disk space required to store \prometheus{} output as a function of the incident neutrino energy.
  Note that the relative disk requirements of each interaction type follow relative fraction of the initial neutrino energy that is deposited in the detector.
  Solid lines correspond to charged-current interactions while dashed lines correspond to neutral current interactions.
  Different colors indicate different neutrino flavors.
  }
  \label{fig:disk_space}
\end{figure}

As discussed above, \prometheus{} events are stored in the \texttt{Parquet} file format.
Typically, the disk space required ranges between 1~kB and 10~kB per event for incident neutrino energies between $10^{2}$~GeV and $10^{5}$~GeV, see Fig.~\ref{fig:disk_space}.
This means that datasets that have millions of events, as is required for many machine-learning applications, can be stored in $\mathcal{O}$(1~GB) of memory.
The exact value will depend on the energy range being simulated, as well as the initial energy spectrum.

In general, the functional dependence on energy and relative ordering of different interactions in Fig.~\ref{fig:disk_space} align with expectations.
For example, higher energy neutrinos lead to higher light yields, and thus a larger number of photons that must be stored.
Furthermore, the relative ordering of the lines makes sense since the number of photons produced should be proportional to the energy deposited in the detector.
This is the trend that is seen in Fig.~\ref{fig:disk_space} at energies above 3~TeV, where storing the photon arrival information drives disk space needs.

It should be noted that not every event produces enough light to trigger the detector.
Our studies have found that the efficiency between injected events and those that trigger a detector is $\sim60$\% for $\nu_{e}$ charged-current interactions and $\sim20$\% for $\nu_{\mu}$ charged-current interactions.
This will depend on the trigger criteria, initial energy spectrum, interaction type, and generation specifications and we postpone detailed study of this for a future work.

%% file: sections/checks.tex
\section{Validation and Unit Tests}
\label{sec:checks}

Almost all the packages we use to model physical processes are published, and as such, have been well-verified.
The only exceptions are the \texttt{fennel} and \texttt{hyperion} packages, which provide a new implementation of the parametrization that has already been shown to work in IceCube simulation~\cite{Wiebusch:thesis,Radel:2012ij} and of ray-tracing-based photon propagation.

\begin{figure}[b]
  \centering
  \includegraphics[width=\textwidth]{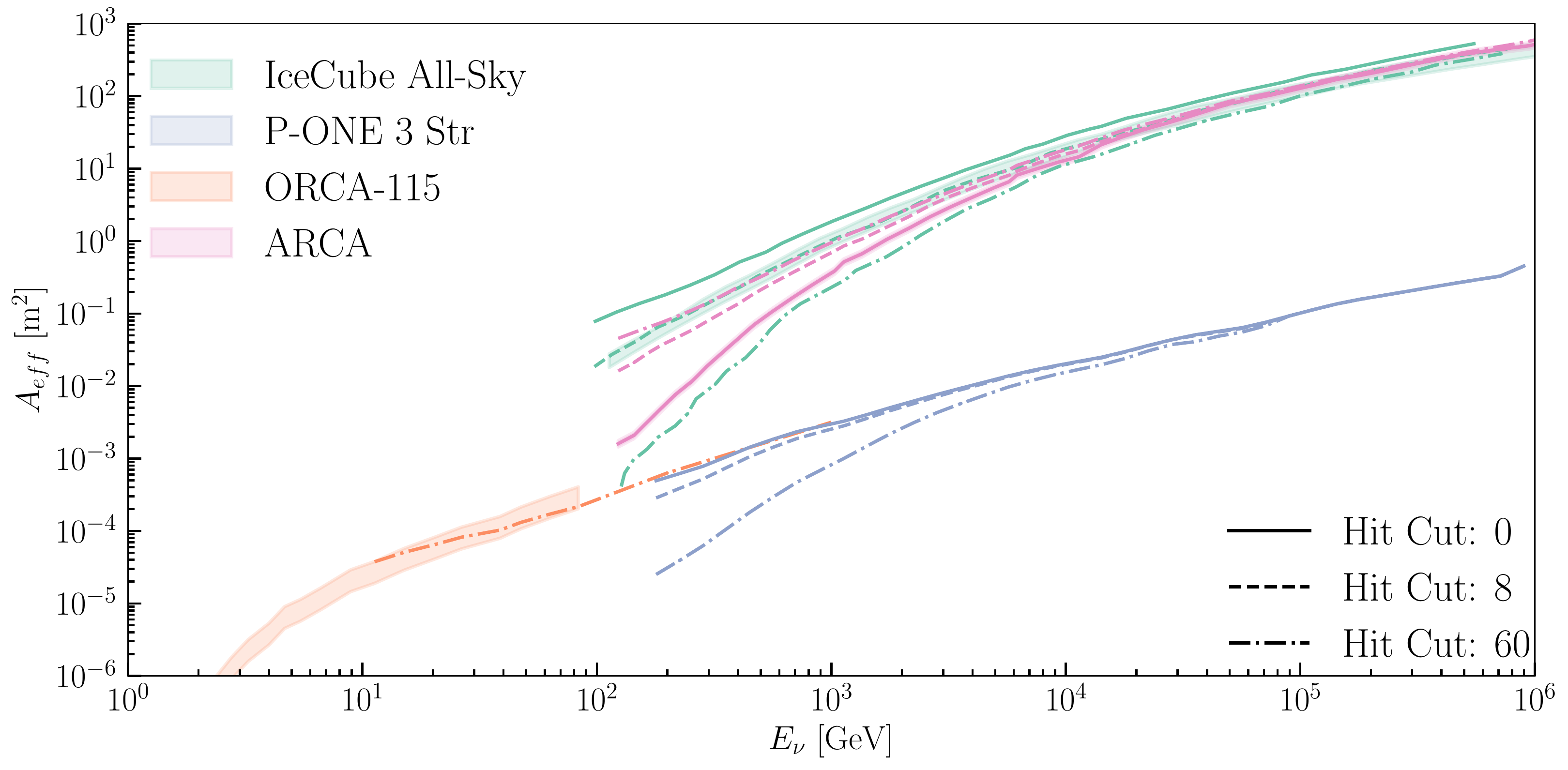}
  \caption{\textbf{\textit{Effective area computed using \prometheus{} with comparisons to published results.}}
  We compare the $\nu_{\mu}$ effective areas computed with \prometheus{} for IceCube, P-ONE3, ORCA, and ARCA for three different hit requirements, denoted by different line styles, to published effective areas.
  The IceCube effective area, taken from~\cite{karle:2009abc}, is for $\nu_{\mu}+\nu_{\tau}$ events which pass the SMT-8 trigger and agrees with our calculation to within uncertainties.
  The ARCA \cite{KM3Net:2016zxf} and ORCA \cite{deWasseige:2020dnq} cases effective areas are constructed with more complicated hit requirements.
  Still, the scale and shape of the ORCA and ARCA effective areas and the \prometheus{} effective areas agree within uncertainties despite the simplified selection criterion.
  As of the publication of this paper, there is no published effective area for P-ONE3.
  }
  \label{fig:effa}
\end{figure}

We must then show that all these packages are working in concert to produce physically meaningful simulation.
Since the effective area is primarily governed by the physics implemented in \prometheus{}---the neutrino-nucleon cross section, lepton range, and photon propagation---and requires that all steps in the simulation are functioning together, it offers a good check of our code.
In Fig.~\ref{fig:effa}, we reproduce published effective areas of several water- and ice-based neutrino telescopes 
It is important to note that this effective area is always defined after some level of cuts, and while we have done our best to reproduce the cuts from the references, there is not always sufficient information to do so perfectly.
Furthermore, the OM response cannot be incorporated without access to proprietary information, and thus we expect differences of $\mathcal{O}(10\%)$.
Users can extend, if needed, our simulation to incorporate these effects for a more detailed comparison.

We are currently working to implement a test suite to ensure the long-term reliability of \prometheus{}.
While this is straightforward for most sections of the code, those that are non-deterministic due to the lack of a seeding option require more careful consideration.
We are working towards statistical tests that when repeatedly applied result in a high level of confidence that the code is performing as expected.

%% file: sections/conclusions.tex
\section{Conclusions and Future Opportunities}
\label{sec:conclusions}

In this paper, we have presented \prometheus{}, an open-source simulation for neutrino telescopes.
It allows the user to simulate ice- and water-based detectors, allowing for arbitrary detector geometries and a variety of injection inputs.
The output of the \prometheus{} package is a \texttt{parquet} file containing true photon arrival quantities and initial event properties.
We have provided two examples to simulate events in example ice- and water-based detectors.
Furthermore, we have benchmarked the runtime performance for both ice- and water-based simulations.

We hope that \prometheus{} will foster a spirit of collaboration in the growing network of neutrino telescopes.
To this end, we intend to make public the large data sets that we have simulated for various detector geometries.
In the meantime, we are happy to share these sets with anyone who wishes to experiment with them and encourage anyone interested to reach out to the authors.
While these datasets should suffice for many use cases, we have set up a form to take simulation requests from the community to cover needs we had not foreseen.
Once again, please reach out to the authors if you are interested in this option.




Through these efforts, we hope that the community will be empowered to design new reconstruction algorithms and that these techniques may be adapted across detectors.
To this end, we have made a \texttt{community\_contributions} directory in the GitHub repository dedicated to collecting community contributions, and we encourage any users that may develop useful tools to link their repositories there.
Currently, this contains two such contributions: a machine-learning-based directional and energy reconstruction capable of running at speeds comparable to the neutrino telescope trigger rates~\cite{Yu:2023ehc}, and a machine-learning algorithm for differentiating single- and dimuon events.

Furthermore, in order to better validate this code, we welcome collaboration with neutrino telescopes.
Ensuring that effective areas match is a good cross check, but we hope that we will be able to compare \prometheus{} to internal simulation on an event-by-event basis.
These collaborations will not only help validate \prometheus{} but also will benefit the internal simulation software by providing independent cross-checks to ensure consistency.
In undertaking these collaborative efforts, we hope we will move one step closer to the consistent, globally meaningful simulation of all experiments.

%% file: sections/acknowledge.tex
\section*{Acknowledgements}
\begin{CJK*}{UTF8}{gbsn}
We would like to thank all users who tested early versions of this software, including---in no particular order---Miaochen Jin (靳淼辰), Eliot Genton, Tong Zhu (朱彤), Rasmus {\O}rs{\o}e, Savanna Coffel, and Felix Yu.
Additionally, we would also like to thank the authors of \texttt{LeptonInjector} \texttt{LeptonWeighter}, \texttt{PPC}, \texttt{PROPOSAL}, and the generations of developers and researchers who laid the groundwork for this package.
In particular, we will like to thank Dima Chirkin for his support in unraveling the mysteries of \texttt{PPC}.

JL and CAA were supported by the Faculty of Arts and Sciences of Harvard University, and the Alfred P. Sloan Foundation through this work.
JL is supported by the NSF under grants PLR-1600823 and PHY-1607644 and by the University of Wisconsin Research Council with funds granted by the Wisconsin Alumni Research Foundation.
DK acknowledges the support of Lynne Sacks and Paul Kim for supporting his research stage at Harvard on the summer of 2022.
SG was supported by the Harvard College Research Program (HRCP) and the PRISE program at Harvard to develop this work.
SMB was supported by the Australian Government through the Australian Research Council's Discovery Projects funding scheme (project DP220101727).
This research was supported by The University of Melbourne’s Research Computing Services and the Petascale Campus Initiative.
\end{CJK*}

%% file: appendices/config.tex
\section{Configuration Details}
\label{app:config}

In this appendix, we enumerate all the options available in the configuration file.
The first column is the name that we have used in the text to refer to these variables, the second is a description of the variables, the third is the path of the variable in the configuration file, and the last is the default value.
Many options are \texttt{None} by default.
If this case, the options can often be inferred from other required options, or have a reasonable default behavior.
A good example of this is cross-section files.
We provide splines of cross sections for Standard Model interactions, and as such, we can find the correct cross sections for the requested interaction if no files are provided by the user.
We will describe the default behavior of such options in the second column.

When giving the configuration path, we will use the syntax of \texttt{Python} formatted strings.
Furthermore, we assume the variable \texttt{RESOURCES\_DIR} points to the \prometheus{} resources directory at \texttt{/resources/}.
We should also note that in the section discussing \texttt{PPC}, we restrict ourselves to the GPU version of the code; however, there is are equivalent fields for the non-GPU version that can be accessed by eliding \texttt{"\_CUDA"} from the paths.

\begin{center}
\begin{tabularx}{\textwidth}{|| p{40pt} | p{150pt} | p{135pt} | X ||}
\hline

Text name & Description & Configuration path & Default value \\
\hline
\hline

--- &
Version information for reference. The numerical value will change for each release. We do not recommend changing this field. &
\texttt{"general/version"} &
\texttt{github} \\
\hline

run number &
Run number. This will be used to set the random state seed if None is given and will dictate the output file names if none are provided. &
\texttt{run/run number}. &
1337\\
\hline

--- & 
The number of events you wish to simulate &
\texttt{run/nevents} &
10 \\
\hline

--- &
&
\texttt{run/storage prefix} &
\texttt{"./output/"}\\
\hline

--- &

--- &
\texttt{run/outfile} &
\texttt{None}\\
\hline

--- &
Seed to be used for random number generation. If this is \texttt{None} the random number seed and the run number will be the same. &
\texttt{"run/random state seed"} &
\texttt{None} \\
\hline
\hline
\texttt{geo} file &
\texttt{.geo} file to get detector configuration \texttt{geofile} information from. If this is left as \texttt{None}, a \texttt{Detector} object must be passed when \prometheus{} is initialized &
\texttt{"detector/specs file"} &
\texttt{None} \\
\hline
\hline

injector software &
Name of the injection software to be used.&
\texttt{"injection/name"} &
\texttt{"LeptonInjector"} \\
\hline

--- &
Whether to carry out a new injection with \leptoninjector{}. &
\texttt{"injection/LeptonInjector/ inject} &
\texttt{True} \\
\hline

--- &
Install location for \leptoninjector{}. This will be used if \leptoninjector{} is not found in the system \texttt{PYTHONPATH}. The default value is the one used by the Singularity and Docker containers.&
\texttt{"injection/LeptonInjector/ install location"} &
 \texttt{"/opt/LI/ install/ lib/ python3.9/ site-packages"} \\
\hline

--- &
Directory where \leptoninjector{} cross-section tables are stored &
\texttt{"injection/LeptonInjector/ xsec dir"} &
\texttt{"\{RESOURCES\_DIR\}/ cross\_section\_ splines/"}\\
\hline

\end{tabularx}
\end{center}

\pagebreak

\begin{center}
\begin{tabularx}{\textwidth}{|| p{40pt} | p{150pt} | p{135pt} | X ||}
\hline

--- &
File with differential cross-section tables &
\texttt{"injection/LeptonInjector/ diff xsec"} &
\texttt{None} \\
\hline

--- &
File with total cross section tables. &
\texttt{"injection/LeptonInjector/ total xsec"}
& \texttt{None} \\
\hline

--- &
File with Earth model parameterization &
\texttt{"injection/LeptonInjector/ earth model location"} &
\texttt{None} \\
\hline

--- &
\texttt{h5} file with \leptoninjector{} injection &
\texttt{"injection/LeptonInjector/ injection file"} &
\texttt{None} \\
\hline

\texttt{lic} file &
\leptoninjector{} configuration file. Necessary for weighting &
\texttt{"injection/LeptonInjector/ lic file"} &
\texttt{None} \\
\hline

\hline
--- &
Where to store \leptoninjector{} output \texttt{HDF5} file. If \texttt{None}, this will be placed in the default output location with a name that is unique for each run number. If the option to create a new injection is turned off, this file must exist & 
\texttt{"injection/LeptonInjector/ injection file"} & 
\texttt{None} \\
\hline

\texttt{lic} file &
Where to store \leptoninjector{} configuration file. If \texttt{None}, this will be saved in the default output location with a unique name for each run number. &
\texttt{"injection/LeptonInjector/ lic file"} &
\texttt{None} \\
\hline

--- &
First of two final states needed to run \leptoninjector{} &
\texttt{"injection/LeptonInjector/ final state 1"} &
\texttt{MuMinus} \\
\hline

--- &
Second of two final states needed to run \leptoninjector{} &
\texttt{"injection/LeptonInjector/ final state 2"} & 
\texttt{Hadrons} \\
\hline

$E_{\rm{min}}$ &
Minimum energy when sampling initial neutrino energies.&
\texttt{"injection/LeptonInjector/ minimal energy"} &
$10^{2}$~GeV\\
\hline

$E_{\rm{max}}$ &
Maximum energy when sampling initial neutrino energies &
\texttt{"injection/LeptonInjector/ maximal energy"}&
$10^{6}$~GeV\\
\hline

$\gamma$ &
The spectral index of the power law to sample energies from. The default value is chosen since this will result in a uniform number of events per bin when using log-spaced bins. Note that there is an implicit negative sign, \textit{i.e.} the event are chosen proportional to $E^{-\gamma}$ &
\texttt{"injection/LeptonInjector/ power law"} &
1\\
\hline

$\theta_{\rm{min}}$ &
Minimum zenith when sampling initial neutrino directions. A value of 0 corresponds to up-going neutrinos. &
\texttt{"injection/LeptonInjector/ min zenith"} &
$0$\\
\hline

$\theta_{\rm{max}}$ &
Maximum zenith when sampling initial neutrino directions. A value of $\pi$ corresonds to down-going neutrinos.&
&
$\pi$ \\
\hline

$\phi_{\rm{min}}$ &
Minimum azimuth when sampling initial neutrino directions. &
& $0$ \\
\hline

\end{tabularx}
\end{center}
\pagebreak

\begin{center}
\begin{tabularx}{\textwidth}{|| p{40pt} | p{150pt} | p{135pt} | X ||}
\hline

$\phi_{\rm{max}}$ &
Maximum azimuth when sampling initial neutrino directions. &
&
$2\pi$\\

\hline

--- &
Whether to do ranged injection. If this is left as \texttt{None}, we will use ranged injection for $\nu_{\mu}$ CC interactions and volume for anything else &
\texttt{"injection/LeptonInjector/ simulation/is ranged"} &
\texttt{None} \\
\hline

$r_{\rm{inj}}$ & \leptoninjector{} injection radius. If this is left as \texttt{None}, we will select a radius that encompasses the whole detector with a padding equal to a absorption length of the medium. & \texttt{"injection/LeptonInjector/ injection radius"} & \texttt{None} \\
\hline

$\ell_{\rm{ec}}$ & \leptoninjector{} endcap length. If this is left as \texttt{None}, we will select a length that encompasses the whole detector with a padding equal to a absorption length of the medium. & \texttt{"injection/LeptonInjector/ endcap lenght"} & \texttt{None} \\
\hline

$r_{\rm{cyl}}$ & \leptoninjector{} cylinder radius. If this is left as \texttt{None}, we will select a radius that encompasses the whole detector with a padding equal to a absorption length of the medium. &
\texttt{"injection/LeptonInjector/ cylinder radius"} &
\texttt{None} \\
\hline

$h_{\rm{cyl}}$ & \leptoninjector{} cylinder height. If this is left as \texttt{None}, we will select a height that encompasses the whole detector with a padding equal to a absorption length of the medium. & \texttt{"injection/LeptonInjector/ cylinder height"} & \texttt{None} \\
\hline

--- &
Whether to inject with \prometheus{}. This is not currrently supported, and changing this value will raise an error. &
\texttt{"injection/prometheus/ inject} & 
\texttt{False}\\
\hline

--- &
Where to find output \prometheus{} file to use as an injection. If this is \texttt{None} and injection from \prometheus{} is requested, this will raise an error.&
\texttt{"injection/prometheus/ paths/injection file"}&
\texttt{None} \\
\hline

--- & 
Whether to inject with \texttt{GENIE}. This is not supported, and changing this  value will raise an error. & \texttt{"injection/GENIE/} & 
\texttt{False}\\
\hline

--- &
Where to find \texttt{GENIE} output injection file. If \texttt{None} and injection from  \texttt{GENIE} is requested, this raise an error since we cannot currently generate new \texttt{GENIE} injection. &
&
\texttt{None} \\
\hline

--- &
Where to find \texttt{GENIE} output \texttt{parquet} file. If \texttt{None}, this raise an error since it cannot be created. &
&
\texttt{None}\\
\hline

\end{tabularx}
\end{center}
\pagebreak

\begin{center}
\begin{tabularx}{\textwidth}{|| p{40pt} | p{150pt} | p{135pt} | X ||}
\hline

Lepton propagator &
Which lepton propagator to use. Currently the only options are \texttt{"new proposal"}, which uses \texttt{PROPOSAL} version \texttt{7.x.x}, and \texttt{"old proposal"} which uses \texttt{PROPOSAL} version \texttt{6.1.6}. As mentioned in the text, this latter option exists for legacy reasons, and we strongly encourage users to use the latest version of \texttt{PROPOSAL}. &
\texttt{"lepton propagator/name"} &
\texttt{"new proposal"} \\

\hline
--- & 
Where to store the tables that \texttt{PROPOSAL} generates. by default they go into a directory in the \texttt{RESOURCE\_DIR} &
\texttt{"lepton propagator/paths/ tables path"} &
"\texttt{\{RESOURCES\_DIR\}/ PROPOSAL\_tables/"} \\
\hline

Earth model&
Which Earth model to use. If not specified, this will use a detector specific Earth model, if one exists, or a generic one corresponding to the medium in which the detector is deployed.&
\texttt{"lepton propagator/paths/ earth model location"}&
\texttt{None} \\
\hline

--- &
Fractional energy cutoff for \texttt{PROPOSAL}. &
\texttt{"lepton propagator/simulation/ vcut"} &
0.1 \\
\hline

--- &
Absolute energy cutoff for \texttt{PROPOSAL}.&
\texttt{"lepton propagator/simulation/ ecut"} &
0.5 GeV\\
\hline

--- &
Whether to track energy losses that don't produce Cherenkov emission&
\texttt{"lepton propagator/simulation/ soft losses"} &
\texttt{False}\\
\hline

--- &
Whether to consider the Landau–Pomeranchuk–Migdal effect. &
\texttt{"lepton propagator/simulation/ lpm effect"} &
\texttt{True}\\
\hline

--- &
Whether to do continuous randomization&
\texttt{"lepton propagator/simulation/ continuous randomization"} &
\texttt{True}\\
\hline

--- &
Whether to use interpolation with PROPOSAL &
\texttt{"lepton propagator/simulation/ interpolate"} &
\texttt{True}\\
\hline

--- &
Scattering model for lepton&
\texttt{"lepton propagator/simulation/ scattering model"} &
\texttt{"Moliere"}\\
\hline

--- &
Medium in which the detector is embedded. If left as \texttt{None}, we will retrieve this information from the \texttt{Detector} object. &
\texttt{"lepton propagator/simulation/ medium"} &
\texttt{None}\\
\hline

Photon propagation software &
Which photon propagation software to use. This should be \texttt{"PPC"}, \texttt{"PPC\_CUDA"}, or \texttt{"olympus"}. If this is left as \texttt{None}, we will set it based on the detector medium, \texttt{PPC} for ice-based detectors and \texttt{"olympus"} for water-based detectors. &
\texttt{"photon propagator/name"} &
\texttt{None} \\
\hline

\end{tabularx}
\end{center}
\pagebreak

\begin{center}
\begin{tabularx}{\textwidth}{|| p{40pt} | p{150pt} | p{135pt} | X ||}
\hline

Photon field name &
What to call the field in the \texttt{Parquet} file that stores photon information. This is an options for legacy compatibility reasons. &
\texttt{"photon propagator/photon field name"}&
\\
\hline

---&
Where the \texttt{olympus} resources are stored. This includes the normalizing flows that are used as well as some configuration information. &
\texttt{"photon propagator/olympus/paths/ location"} &
\texttt{"\{RESOURCES\_DIR\}/ olympus\_resources"}\\
\hline

--- &
\texttt{olympus} photon model. Defines optical medium properties. These will be used for scattering and attenuation. &
\texttt{"photon propagator/olympus/paths/ photon model"} &
\texttt{pone\_config.json}\\
\hline

---&
\texttt{olympus} timing normalizing flow. These distributions are sampled to generate the time distributions. &
\texttt{"photon propagator/olympus/paths/ flow"} &
\texttt{"photon\_arrival\_ time\_nflow\_ params.pickle"}\\
\hline

---&
\texttt{olympus} counts. These Distributions are sampled to generate the number of hits. &
\texttt{"photon propagator/olympus/paths/ counts"} &
\texttt{"photon\_arrival\_ time\_counts\_ params.pickle"}\\
\hline

--- & BETA: Generates distribution tables on the fly. Currently not supported.
&
\texttt{"photon propagator/olympus/ simulation/files"}&
\texttt{True}\\
\hline

--- & Wavelength in nm of the photons to generate. 700 nm are used as a benchmark value. 420 nm are recommeded for production.
&
\texttt{"photon propagator/olympus/ simulation/wavelength"}&
700\\
\hline

--- & Splits the detector into chuncks of size 'splitter'. Smaller sizes reduce the memory usage, while increasing simulation time.
&
\texttt{"photon propagator/olympus/ simulation/splitter"} &
10000\\
\hline

--- &
Directory with the \texttt{PPC} executable compiled for a GPU. &
\texttt{"photon propagator/PPC\_CUDA/paths/ location"} &
\texttt{"{RESOURCES\_DIR}/ PPC\_executables/ PPC\_CUDA"}\\
\hline

--- &
Temporary directory where tables for \texttt{PPC} will be put while the program is running. If this path exists, the program will error unless \texttt{force} is set to true. Please note, that if you are running parallel jobs, this directory should be for each job to avoid a race condition.&
\texttt{"photon propagator/PPC\_CUDA/paths/ ppc\_tmpdir"}&
\texttt{./.ppc\_tmpdir}\\
\hline

--- &
If selected, the program will not error if the \texttt{ppc\_tmpdir} exists, and instead will remove it, and put a blank directory in its place.&
\texttt{"photon propagator/PPC\_CUDA/paths/ force"} &
False \\
\hline

--- &
Location of intermediate file where OM hits from \texttt{PPC} will be stored. &
\texttt{"photon propagator/PPC\_CUDA/paths/ ppc\_tmpfile"}&
\texttt{".event\_ hits.ppc.tmp"}\\
\hline

--- &
Location of intermediate file where energy losses will be stored. &
\texttt{"photon propagator/PPC\_CUDA/paths/ f2k\_tmpfile"}&
\texttt{".event\_ losses.f2k.tmp"}\\
\hline

\end{tabularx}
\end{center}
\pagebreak

\begin{center}
\begin{tabularx}{\textwidth}{|| p{40pt} | p{150pt} | p{135pt} | X ||}
\hline

Tables directory &
Directory where the tables that \texttt{PPC} needs are located. By default, we give the path to a directory that has a parameterization of the South Pole ice, and a parameterization of the OM angular acceptance that accepts all photons. &
\texttt{"photon propagator/PPC\_CUDA/paths/ ppctables"}&
\texttt{"{RESOURCES\_DIR}/ PPC\_tables/ ic\_accept\_all/"}\\
\hline

--- &
Which GPU to run on. &
\texttt{"photon propagator/PPC\_CUDA/ simulation/ device"}&
0 \\
\hline

--- &
Whether to suppress the large amount of output that \texttt{PPC} logs via standard error. This can be helpful for debugging.&
\texttt{"photon propagator/PPC\_CUDA/ simulation/ suppress\_output"}&
True \\
\hline

\end{tabularx}
\end{center}

%% file: appendices/examples.tex
\section{Examples}
\label{app:examples}

In this appendix, we offer a more comprehensive view of some of the options available to the user.
We will try to point out why we are making certain decisions or setting specific options when relevant.
Many of these examples have a corresponding \texttt{Jupyter} notebook in the \texttt{/examples/} directory.
These include example plotting for the interactive environment.
If a particular example has a notebook, the section header will link to it.

\subsection{\href{https://github.com/Harvard-Neutrino/prometheus/blob/main/examples/output_example.ipynb}{Examining the output}}
\label{ex:output}

We will first take a look at the \prometheus{} output, primarily focusing on what information is contained in the file.
As mentioned in the main text, the files are stored in the \texttt{Parquet} format; however, all examples here will use the \texttt{Awkward} package to read these.
While we will not showcase any here, the \texttt{Awkward} package has many useful features; see this \href{https://www.youtube.com/watch?v=WlnUF3LRBj4&t=1039s&pp=ygUhYXdrd2FyZCBhcnJheSBoaWdoIGVuZXJneSBwaHlzaWNz}{presentation} for some examples.

We will begin by importing \texttt{Awkward} and \texttt{numpy}

\begin{lstlisting}[language=python]
import awkward as ak
import numpy as np
\end{lstlisting}

Next we will load in an example file that is provided in the GitHub repository.
The exact nature of this file will be explained throughout the example, so we will refrain from explaining it here.

\begin{lstlisting}[language=python, firstnumber=3]
events = ak.from_parquet("./output/example\_photons.parquet")    
\end{lstlisting}

One quantity that we can compute straightforwardly is the number of events that this file contains.
To do this, we can use \texttt{Python}'s built-in \texttt{len} function.

\begin{lstlisting}[language=python, firstnumber=4]
print(f"Number of events simulated: {len(events)}")
\end{lstlisting}

For now will turn our attention to just one event since most of the relevant features can be demonstrated with this simplified case.

\begin{lstlisting}[language=python, firstnumber=5]
event = events[7]
\end{lstlisting}

It should be noted, though, that most information for all events in the file can be accessed by change \texttt{event} to \texttt{events}.

Let us start by looking into the Monte Carlo truth information, which resides in the \texttt{mc\_truth} field.
A good summary of it can be displayed in a \texttt{Jupyter}
notebook by accessing it on the last line.

\begin{lstlisting}[language=python, firstnumber=6]
event["mc_truth"]
\end{lstlisting}

We may wish to know about the type of particle that produced this event.
We can access relevant quantities about in the fields that start with \texttt{initial\_state\_*}.
We can list all of these by running that following line.

\begin{lstlisting}[language=python, firstnumber=7]
f"We can look at {[x for x in event['mc_truth'].fields if 'initial' in x]}"
\end{lstlisting}

Let's take a look at the initial particle energy and type.

\begin{lstlisting}[language=python, firstnumber=8]
init_type = event["mc_truth", "initial_state_type"]
init_e = event["mc_truth", "initial_state_energy"]
print(f"This initial particle was a {init_type} with energy {init_e} GeV.")
\end{lstlisting}

Furthermore, we can look at the interaction vertex.

\begin{lstlisting}[language=python, firstnumber=11]
init_vertex = np.array([
    event["mc_truth", "initial_state_x"],
    event["mc_truth", "initial_state_y"],
    event["mc_truth", "initial_state_z"]
])
print(f"The interaction vertex was at {init\_vertex} m")
\end{lstlisting}

We can look deeper into the initial state variables, but let us now turn our attention to the final state variables.
We can see which fields are available to use with the following line.

\begin{lstlisting}[language=python, firstnumber=17]
f"We can look at {[x for x in event['mc_truth'].fields if 'final' in x]}"
\end{lstlisting}

It should be noted here that all of these fields contain lists, since we cannot know the number of final state particles \textit{a priori}.

As an example of this, let's examine the particles that were produced from the initial interaction.

\begin{lstlisting}[language=python, firstnumber=18]
final_type = event["mc_truth", "final_state_type"]
print(f"The final products of this interaction are {[x for x in final_type]}")
\end{lstlisting}

Now we can look at the energy of these particles with the following line.

\begin{lstlisting}[language=python, firstnumber=21]
final_e = event["mc_truth", "final_state_energy"]
print(f"The final particles had energies {final_e} GeV")
\end{lstlisting}

Now that we have a reasonable understanding of the \texttt{mc\_truth} information, let us look into the photons that arrived at OMs.

First, we can see what fields are avaiable to use with the following line.

\begin{lstlisting}[language=python, firstnumber=24]
event["photons"].fields
\end{lstlisting}

We can then look at the number of photons the arrived at OMs and see how many unique OMs saw light.

\begin{lstlisting}[language=python, firstnumber=25]
print(f"The first event produced {len(event['photons', 't'])} photons that reached an OM")

unique_om = list(set(x for x in zip(event["photons", "string_id"], event["photons", "sensor_id"])))
print(f"The number of OMs that saw light is {len(unique_om)}")
\end{lstlisting}

We can then get a sense of the timing distribution by looking at the \texttt{"t"} field.

\begin{lstlisting}[language=python, firstnumber=29]
times = event["photons", "t"]
print(f"The first photon arrived at {np.min(times)} ns and the last one arrived at {np.max(times)} ns")
\end{lstlisting}

Finally, we can use the \texttt{id\_idx} field to find which final state particle produced the photon.
The $-1$ that appears here is conventional.
A value of $0$ corresponds to the initial neutrino, which is not included in the final products list.

\begin{lstlisting}[language=python, firstnumber=31]
which_final = event["mc_truth", "final_state_type", event["photons", "id_idx"]-1]
print([pdg_dict[x] for x in which_final])
\end{lstlisting}

By accessing the metadata of the \texttt{Parquet} file, we can see the configuration \texttt{dict} that produced this simulation.
A couple additional libraries are needed to access this information.
Assuming these are installed, you may run the following code.

\begin{lstlisting}[language=python, firstnumber=33]
import pyarrow.parquet as pq
import json

config = json.loads(pq.read_metadata('./output/example_photons.parquet').metadata[b'config_prometheus'])
for k, v in config.items():
    print(k)
    print(v)
    print()
\end{lstlisting}

\subsection{\href{https://github.com/Harvard-Neutrino/prometheus/blob/mainx/examples/detector_example.ipynb}{Building a detector}}
\label{ex:detector_build}

In this section, we will show how to use the utilities from \prometheus{} to construct a detector, including building a new detector and reading one in from a \texttt{geo} file.

Let's start by building a detector in ice!
First we import the a utility from the \texttt{detector\_factory}, as well as the \prometheus{} \texttt{Medium} object.
This detector will lie on a orthogonal grid.

\begin{lstlisting}[language=python]
from prometheus.detector import Medium
from prometheus.detector.detector_factory import make_grid
\end{lstlisting}

Next, we will specify the parameters of the detector, and construct it.

\begin{lstlisting}[language=python, firstnumber=3]
n_per_side = 9 # Number of strings per side
string_spacing = 120 # Spacing in meters between strings
oms_per_string = 60 # Number of OMs per string
om_spacing = 15 # Distance between OMs on same string in meters
z_cent = -2000 # z-coordinate of the center of the detector
medium = Medium.ICE # Medium in which the detector is embedded

ice_det = make_grid(
    n_per_side,
    string_spacing,
    oms_per_string,
    oms_per_string,
    z_cent,
    medium
)
\end{lstlisting}

We now have a detector!
Let's write it to a \texttt{geo} file for later use.

\begin{lstlisting}[language=python, firstnumber=18]
import prometheus
resource_dir = f"{'/'.join(prometheus.__path__[0].split ( '/ ')[: -1]) }/resources/"

ice_det.to_geo(f"{resources_dir}/geofiles/demo_ice.geo")
\end{lstlisting}

Easy.
Now let's load it back up and make sure we actually have the same detector.

\begin{lstlisting}[language=python, firstnumber=22]
from prometheus.detector import detector_from_geo

det2 = detector_from_geo(f"{resources_dir}/geofiles/demo_ice.geo")

keys = [module.key for module in ice_det.modules]
keys2 = [module.key for module in det2.modules]

matched = True
for key in keys:
    if not all(ice_det[key].pos==det2[key].pos):
        matched = False
        
print(f"All the keys from the original detector match the saved version: {matched}")

matched = True
for key in keys2:
    if not all(ice_det[key].pos==det2[key].pos):
        matched = False
print(f"All the keys from the saved detector match the original version: {matched}")

print(f"The media are the same: {ice_det.medium==det2.medium}")
\end{lstlisting}

Hopefully that all worked and we can move on to constructing a detector in water.
We will construct this one on a hexagonal grid, and create a denser subdetector.
First, we will construct the larger portion of the detector.

\begin{lstlisting}[language=python, firstnumber=43]
from prometheus.detector import make_hex_grid

n_per_side = 6 # Number of strings per side
string_spacing = 120 # Spacing in meters between strings
oms_per_string = 60 # Number of OMs per string
om_spacing = 15 # Distance between OMs on same string in meters
z_cent = -2500 # z-coordinate of the center of the detector
medium = Medium.WATER # Medium in which the detector is embedded

water_det = make_hex_grid(
    n_per_side,
    string_spacing,
    oms_per_string,
    om_spacing,
    z_cent,
    medium
)
\end{lstlisting}

Now, we will make the denser subregion.

\begin{lstlisting}[language=python, firstnumber=60]
n_per_side = 3 # Number of strings per side
string_spacing = 30 # Spacing in meters between strings
oms_per_string = 60 # Number of OMs per string
om_spacing = 5 # Distance between OMs on same string in meters
z_cent = -2200 # z-coordinate of the center of the detector
medium = Medium.WATER # Medium in which the detector is embedded

sub_det = make_hex_grid(
    n_per_side,
    string_spacing,
    oms_per_string,
    om_spacing,
    z_cent,
    medium
)
\end{lstlisting}

We can now combine these using the built in \texttt{Python} \texttt{+} operator.
We will also save it to a file for later use.

\begin{lstlisting}[language=python, firstnumber=75]
full_det = water_det + sub_det
full_det.to_geo(f"{resource_dir}/geofiles/demo_water.geo")
\end{lstlisting}

If you have \texttt{matplotlib} installed, you can view the detector with the \texttt{display} method of the \texttt{detector}

\begin{lstlisting}[language=python, firstnumber=77]
full_det.display()
full_det.display(elevation_angle=3.14159/2)
\end{lstlisting}

As a note, you cannot combine \texttt{detector}s that are in different media, so the following will throw an error.

\begin{lstlisting}[language=python, firstnumber=79]
mixed_det = ice_det + water_det
\end{lstlisting}

\subsection{\href{https://github.com/Harvard-Neutrino/prometheus/blob/main/examples/injection.ipynb}{Ranged and Volume Injection}}
\label{ex:injection}

We are going to start with a ranged injection of $\nu_{\mu}$ charged current events.
This takes into account the range that the emerging $\mu^{-}$ can travel and chooses the interaction vertex an appropriate distance away.
After that, we will do a volume injection, which selects the interaction vertex in a predefined cylinder near the detector.
Before that we need to import our configuration file.

\begin{lstlisting}[language=python]
import sys
sys.path.append("..")

from prometheus import config
\end{lstlisting}

Now, we need to set a detector.
\prometheus{} relies on the detector being set to select the Earth model and move the injection into physical coordinates.

\begin{lstlisting}[language=python, firstnumber=5]
config["detector"]["geo file"] = "../resources/geofiles/demo_ice.geo"
\end{lstlisting}

For ease of use, let's isolate the \texttt{injection} portion of the configuration and see what options we have available to us.

\begin{lstlisting}[language=python, firstnumber=6]
injection_config = config["injection"]["LeptonInjector"]

import pprint
pprint.pprint(injection_config)
\end{lstlisting}

As mentioned in~\ref{app:config}, all fields which are left \texttt{None} can be set internally based on the values of other fields.
We will leave those alone for now, with the exception of the output paths.

\begin{lstlisting}[language=python, firstnumber=10]
injection_config["paths"]["injection file"] = "./output/cool_new_injection.h5"
injection_config["paths"]["lic file"] = "./output/cool_new_configuration.lic"
\end{lstlisting}

We can now set the simulation parameters that will control how our injection happens.
For this example, we will do an all-sky injection of $\nu_{\mu}$ charged current events, with default energy settings.
This means the energies will range from $10^{2}$~GeV to $10^6$~GeV and will be sampled from a power law with spectral index 1.

\begin{lstlisting}[language=python, firstnumber=12]
# Lots of events
config["run"]["nevents"] = 10_000
# All zenith angle
injection_config["simulation"]["min zenith"] = 0.0
injection_config["simulation"]["max zenith"] = 180.0
# This sets it to numu cc
injection_config["simulation"]["final_1"] = "MuMinus" # "MuPlus" would inject with numubar cc
injection_config["simulation"]["final_2"] = "Hadrons"
\end{lstlisting}

Now we import the \texttt{Prometheus} object, initialize it, and inject.

\begin{lstlisting}[language=python, firstnumber=20]
from prometheus import Prometheus

p = Prometheus(config)
p.inject()
\end{lstlisting}

This may take a minute or two, once it is done though, we can inspect the \texttt{HDF5} \leptoninjector{} output.
First we will check out the energy and directional distributions of the injected events.
We will plot them using appropriate variables and binning so that we expect flat distributions.
This makes checking the injection easy, but does make for visually complex plots, sorry.

\begin{lstlisting}[language=python, firstnumber=24]
import h5py as h5
import numpy as np
import os
import matplotlib.pyplot as plt
from matplotlib.gridspec import GridSpec
plt.style.use(os.path.abspath("../paper_plots/paper.mplstyle"))

h5f = h5.File("./output/cool_new_injection.h5", "r")

fig = plt.figure(constrained_layout=True, figsize=(12,3))
gs = GridSpec(1, 3, figure=fig)
axs = [fig.add_subplot(g) for g in gs]

# Histogram the energies in log-spaced bins
e_edges = np.logspace(2, 6, 17)
e_centers = (e_edges[1:] + e_edges[:-1]) / 2
h0, _ = np.histogram(h5f["RangedInjector0"]["properties"]["totalEnergy"], bins=e_edges)

# Histogram the cosine of the zenith
czen_edges = np.linspace(-1, 1, 17)
czen_centers = (czen_edges[1:] + czen_edges[:-1]) / 2
h1, _ = np.histogram(np.cos(h5f["RangedInjector0"]["properties"]["zenith"]), bins=czen_edges)

# Histogram the azimuth
az_edges = np.linspace(0, 2*np.pi, 17)
az_centers = (az_edges[1:] + az_edges[:-1]) / 2
h2, _ = np.histogram(h5f["RangedInjector0"]["properties"]["azimuth"], bins=az_edges)

hs = [h0, h1, h2]
cents = [e_centers, czen_centers, az_centers]
colors = ["crimson", "dodgerblue", "darkviolet"]
xlabels = [r"$E_{\rm{\nu}}~\left[\rm{GeV}\right]$", r"$\cos\left(\theta_{\rm{zen}}\right)$", r"$\phi_{\rm{az}}$"]

for idx, ax in enumerate(axs):
    ax.step(cents[idx], hs[idx], where="mid", c=colors[idx])
    ax.axhline(10_000 / len(e_centers), label="Expectation", c=colors[idx], ls="--")
    ax.set_xlim(cents[idx][0], cents[idx][-1])
    ax.set_ylim(580, 680)
    ax.set_xlabel(xlabels[idx])
    if idx != 0:
        ax.set_yticklabels([])
    else:
        ax.set_ylabel(r"$N_{\rm{evts}}$")

axs[0].semilogx()
plt.show()
\end{lstlisting}

Hopefully you see a lot of lines fluctuating around the expectation.
Now, let's make a scatter plot of the interaction vertex, with the colors corresponding to the injection zenith angle, $\theta_{\rm{zen}}$.
This will allow us to confirm that the neutrinos are heading towards the detector, and, as a happy byproduct, will give us a plot which is more fun to look at.

\begin{lstlisting}[language=python, firstnumber=71]
fig, ax = plt.subplots(figsize=(20,5))

sct = ax.scatter(
    h5f["RangedInjector0"]["properties"]["x"] / 1000,
    h5f["RangedInjector0"]["properties"]["z"] / 1000,
    c=h5f["RangedInjector0"]["properties"]["zenith"],
    cmap="rainbow",
    alpha=0.5
)
sct2 = ax.scatter([0], [-2], marker="*", label="Detector", c="k")
ax.set_xlim(-300, 300)
ax.set_xlabel(r"$x_{\rm{int}}~\left[\rm{km}\right]$")
ax.set_ylabel(r"$z_{\rm{int}}~\left[\rm{km}\right]$")
cbar = plt.colorbar(sct, label=r"$\theta_{\rm{zen}}$")
ax.legend(loc=2, fontsize=20)
plt.show()
\end{lstlisting}

Nice.
Note that the events with $\theta_{\rm{zen}}\simeq\pi$ are coming from above the detector.
This highlights that the zenith and azimuth angles are defined with respect to the direction of the particle momentum.

Now that we have done that, let's do another injection, this time taking advantage of the volume injection option.
This makes sure that the interaction happens within a predefined cylinder.
This is the default behavior for $\nu_{e}$ charged current and $\nu_{\alpha}$ neutral current interactions; however, we can force this for $\nu_{\mu}$ charged current interactions if, for instance, we wanted to simulate starting $\nu_{\mu}$ charged current events.
First, let's rename the output files so that we don't overwrite things.

\begin{lstlisting}[language=python, firstnumber=87]
injection_config["paths"]["injection file"] = "./output/cool_new_volume_injection.h5"
injection_config["paths"]["lic file"] = "./output/cool_new_volume_configuration.lic"
\end{lstlisting}

We can now set some of the parameters that are \texttt{None} by default, namely the injection cylinder radius and height as well as the \texttt{"is ranged"} flag.
Since these are now set, \prometheus{} will not use the default settings.
You can also uncomment the line which sets the seed which sets the random state seed if you want the injection to be fully independent of the previous injection.

\begin{lstlisting}[language=python, firstnumber=89]
injection_config["simulation"]["is ranged"] = False
injection_config["simulation"]["cylinder radius"] = 700 # m
injection_config["simulation"]["cylinder height"] = 1000 # m
#config["run"]["random state seed"] = 925

p = Prometheus(config)
\end{lstlisting}

We now make the same plots as before to check out this injection.

\begin{lstlisting}[language=python, firstnumber=95]
h5f = h5.File("./output/cool_new_volume_injection.h5", "r")

fig = plt.figure(constrained_layout=True, figsize=(12,3))
gs = GridSpec(1, 3, figure=fig)
axs = [fig.add_subplot(g) for g in gs]

# Histogram the energies in log-spaced bins
e_edges = np.logspace(2, 6, 17)
e_centers = (e_edges[1:] + e_edges[:-1]) / 2
h0, _ = np.histogram(h5f["VolumeInjector0"]["properties"]["totalEnergy"], bins=e_edges)

# Histogram the cosine of the zenith
czen_edges = np.linspace(-1, 1, 17)
czen_centers = (czen_edges[1:] + czen_edges[:-1]) / 2
h1, _ = np.histogram(np.cos(h5f["VolumeInjector0"]["properties"]["zenith"]), bins=czen_edges)

# Histogram the azimuth
az_edges = np.linspace(0, 2*np.pi, 17)
az_centers = (az_edges[1:] + az_edges[:-1]) / 2
h2, _ = np.histogram(h5f["VolumeInjector0"]["properties"]["azimuth"], bins=az_edges)

hs = [h0, h1, h2]
cents = [e_centers, czen_centers, az_centers]
colors = ["crimson", "dodgerblue", "darkviolet"]
xlabels = [r"$E_{\rm{\nu}}~\left[\rm{GeV}\right]$", r"$\cos\left(\theta_{\rm{zen}}\right)$", r"$\phi_{\rm{az}}$"]

for idx, ax in enumerate(axs):
    ax.step(cents[idx], hs[idx], where="mid", c=colors[idx])
    ax.axhline(10_000 / len(e_centers), label="Expectation", c=colors[idx], ls="--")
    ax.set_xlim(cents[idx][0], cents[idx][-1])
    ax.set_ylim(580, 680)
    ax.set_xlabel(xlabels[idx])
    if idx != 0:
        ax.set_yticklabels([])
    else:
        ax.set_ylabel(r"$N_{\rm{evts}}$")

axs[0].semilogx()
plt.show()
\end{lstlisting}

These plots should be identical to those from the previous example if you did not set the random state by hand.
Once again, let's plot the interaction vertex.

\begin{lstlisting}[language=python, firstnumber=134]
fig, ax = plt.subplots(figsize=(7,5))

sct = ax.scatter(
    h5f["VolumeInjector0"]["properties"]["x"] / 1000,
    h5f["VolumeInjector0"]["properties"]["z"] / 1000,
    c=h5f["VolumeInjector0"]["properties"]["zenith"],
    cmap="rainbow",
    alpha=0.5
)
sct2 = ax.scatter([0], [-2], marker="*", label="Detector", c="k")
ax.set_xlim(-0.8, 0.8)
ax.set_ylim(-2.8, -1.2)
ax.set_xlabel(r"$x_{\rm{int}}~\left[\rm{km}\right]$")
ax.set_ylabel(r"$z_{\rm{int}}~\left[\rm{km}\right]$")
cbar = plt.colorbar(sct, label=r"$\theta_{\rm{zen}}$")
ax.legend(loc=2, fontsize=20)
plt.show()
\end{lstlisting}

Note the vastly different scale from the previous plot.
\label{ex:starting_events}

%% file: appendices/earth_models.tex
\section{Earth Models}
\label{app:earth_models}

In this appendix, we show the Earth model's that are used for each detector.
We supply per-detector Earth models because the column depth for down-going neutrinos is significantly impacted by the depth of the medium in which the detector is deployed.
This, in turn, heavily impacts the weighting and thus the expected rate of down-going neutrinos.
Since this is primarily concerned with trajectories where the column depth is dominated by the deployment medium, we use an onion-like model to add a layer of ice or water that is equal to the depth of the local medium.

These plots only differ in the thickness and type of the detector medium, \textit{i.e.} the small sliver of blue ont he left of the plot.
As mentioned above, we use a simplified model of the Earth in \texttt{PROPOSAL}, where we make each layer have a constant density equal to the mean of the density at either end.
To show the validity of this argument, we show the region of lepton propagation---defined as the region within which 99.9\% of charged leptons with energies less than $10^8$~GeV, capable of arriving at the detetor would be contained---in green background.
As one can see, the density is quite flat in this region.
As a note, this region is roughly the same for each detector since the column depth is dominated by the rock component, which does not vary between detectors.

\subsection{ARCA Earth Model}

The water at Capo Passero, where the ARCA detector will be deployed at, has a depth of 3500~m.
Thus, the ARCA earth model is the PREM model with 3500~m of water and 103~km of air above it.

\begin{figure}[h]
  \centering
  \includegraphics[width=0.6\textwidth]{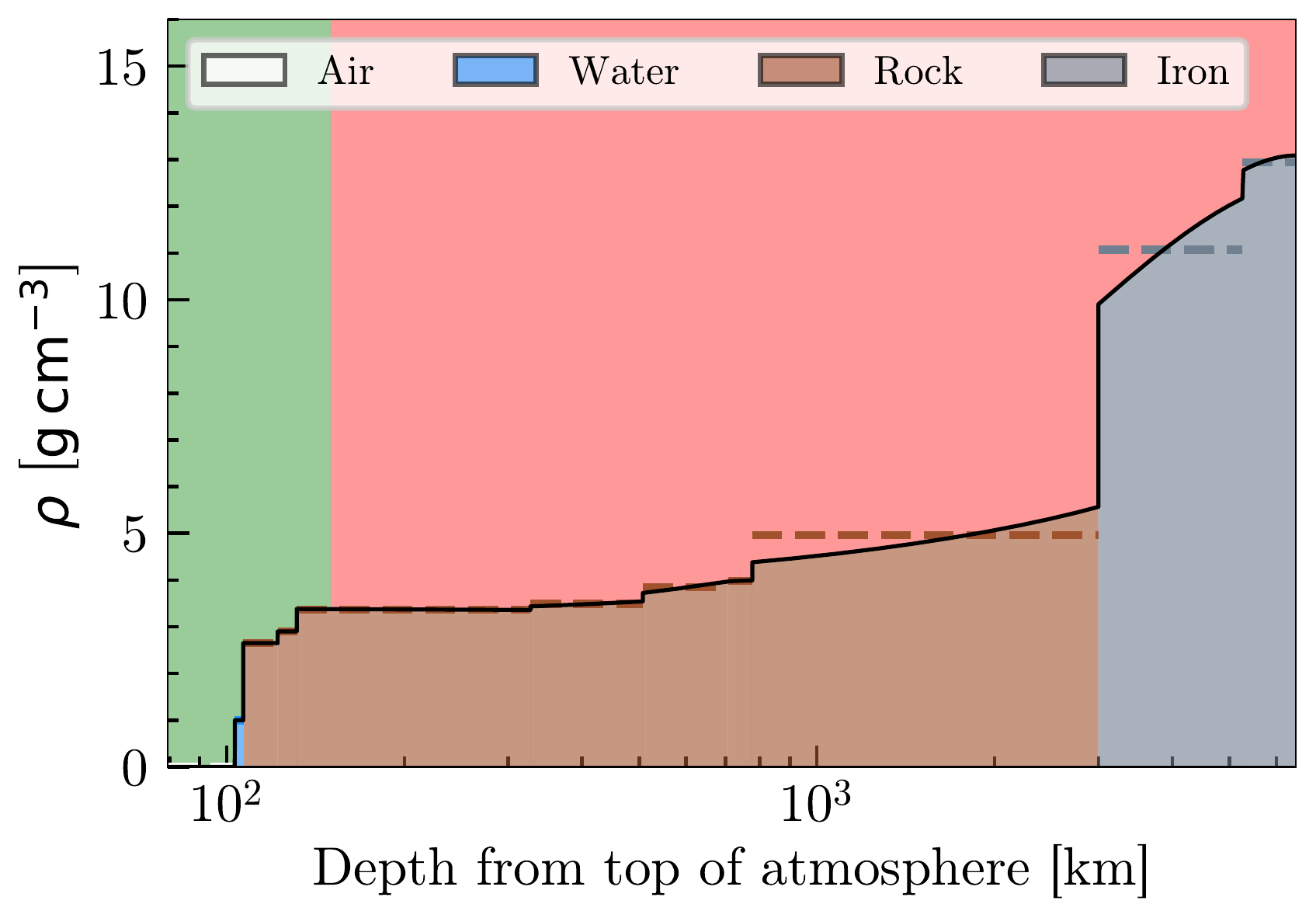}
  \caption{
     \textbf{\textit{Earth model for ARCA}}
  }
  \label{fig:earth_arca}
\end{figure}

\subsection{GVD Earth Model}

While Lake Baikal has a maximum depth of more than 1,600~m, the lake is 1,366~m deep.
Thus the Baikal-GVD Earth model is the PREM model with 1,366~m of water, and 105~km of air above it.

\begin{figure}[h]
  \centering
  \includegraphics[width=0.6\textwidth]{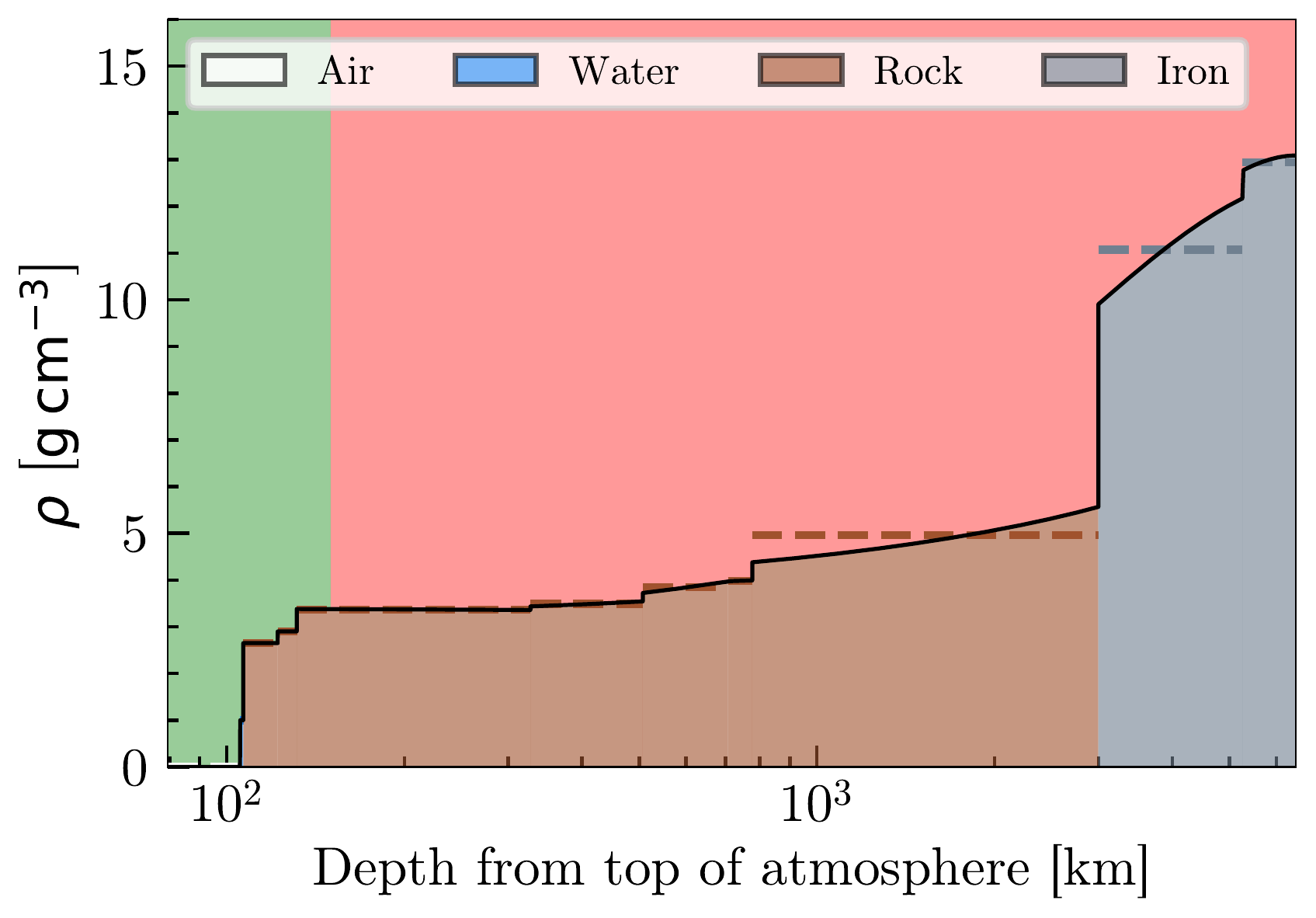}
  \caption{
     \textbf{\textit{Earth model for GVD}}
  }
  \label{fig:earth_gvd}
\end{figure}

\subsection{ORCA Earth Model}

The depth of the Ligurian Sea where the ORCA detector will be deployed is 2,475~m.
Thus the ORCA Earth model is the PREM model with 2,475~m of water, and 104~km of air above it.

\begin{figure}[h]
  \centering
  \includegraphics[width=0.6\textwidth]{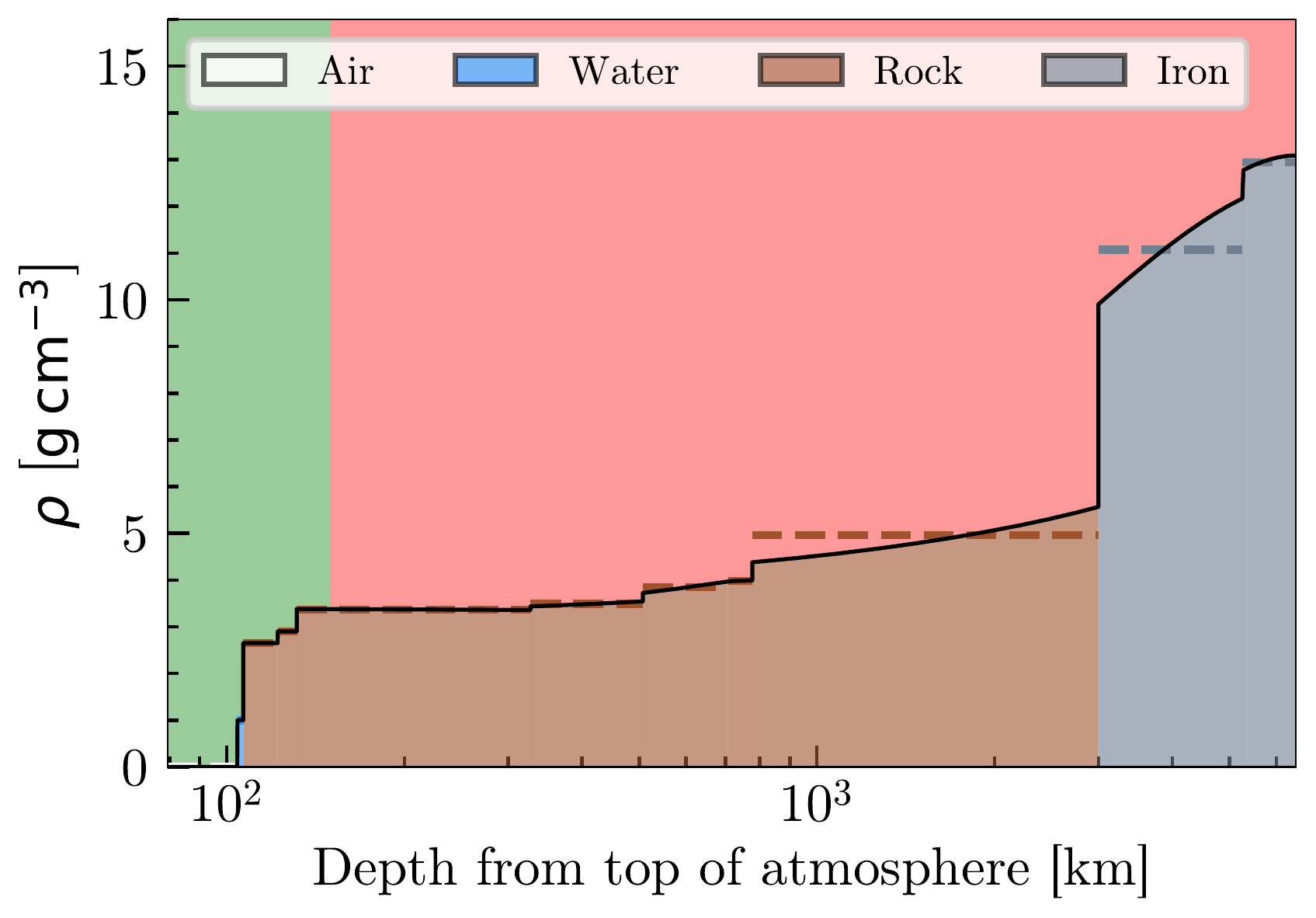}
  \caption{
     \textbf{\textit{Earth model for ORCA}}
  }
  \label{fig:earth_orca}
\end{figure}

\subsection{P-ONE Earth Model}

At the P-ONE deployment site, the Pacific Ocean has a depth of 2,860~m.
Thus the P-ONE Earth model is the PREM model with 2,860~m of water, and 104~km of air above it.

\begin{figure}[h]
  \centering
  \includegraphics[width=0.6\textwidth]{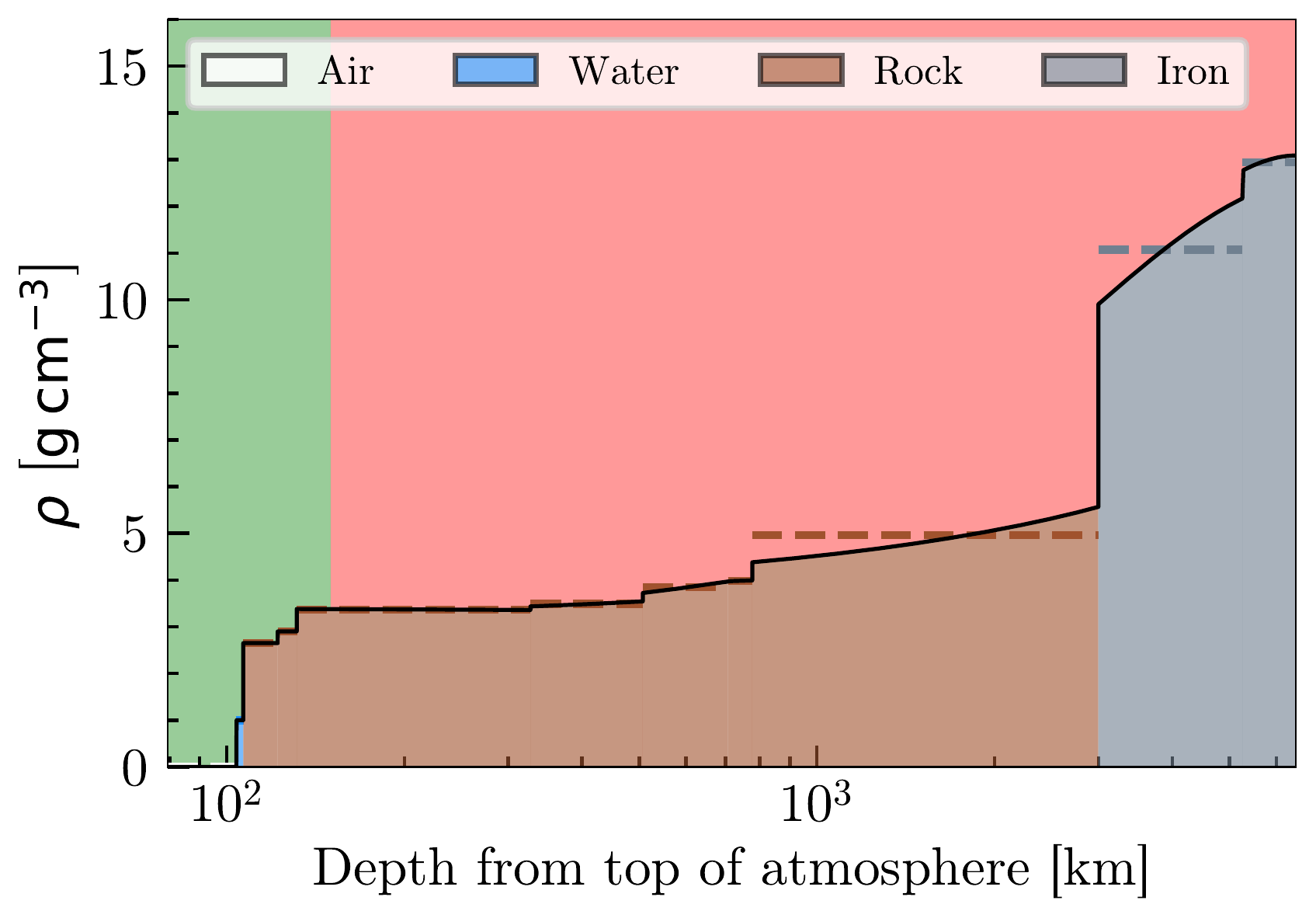}
  \caption{
     \textbf{\textit{Earth model for P-ONE}}
  }
  \label{fig:earth_pone}
\end{figure}

\subsection{South Pole Earth Model}

At the South Pole, there are two layers of frozen water on top of the PREM model.
The layer nearer the rock is the ice in which the instrumented volume lies, and the further layer is compacted snow that has not yet become ``ice.''
These have thicknesses of 2,610~m and 200~m respectively.
These are both accounted for in the South Pole Earth model.
Furthermore, 104~km of air are put above this.

\begin{figure}[h]
  \centering
  \includegraphics[width=0.6\textwidth]{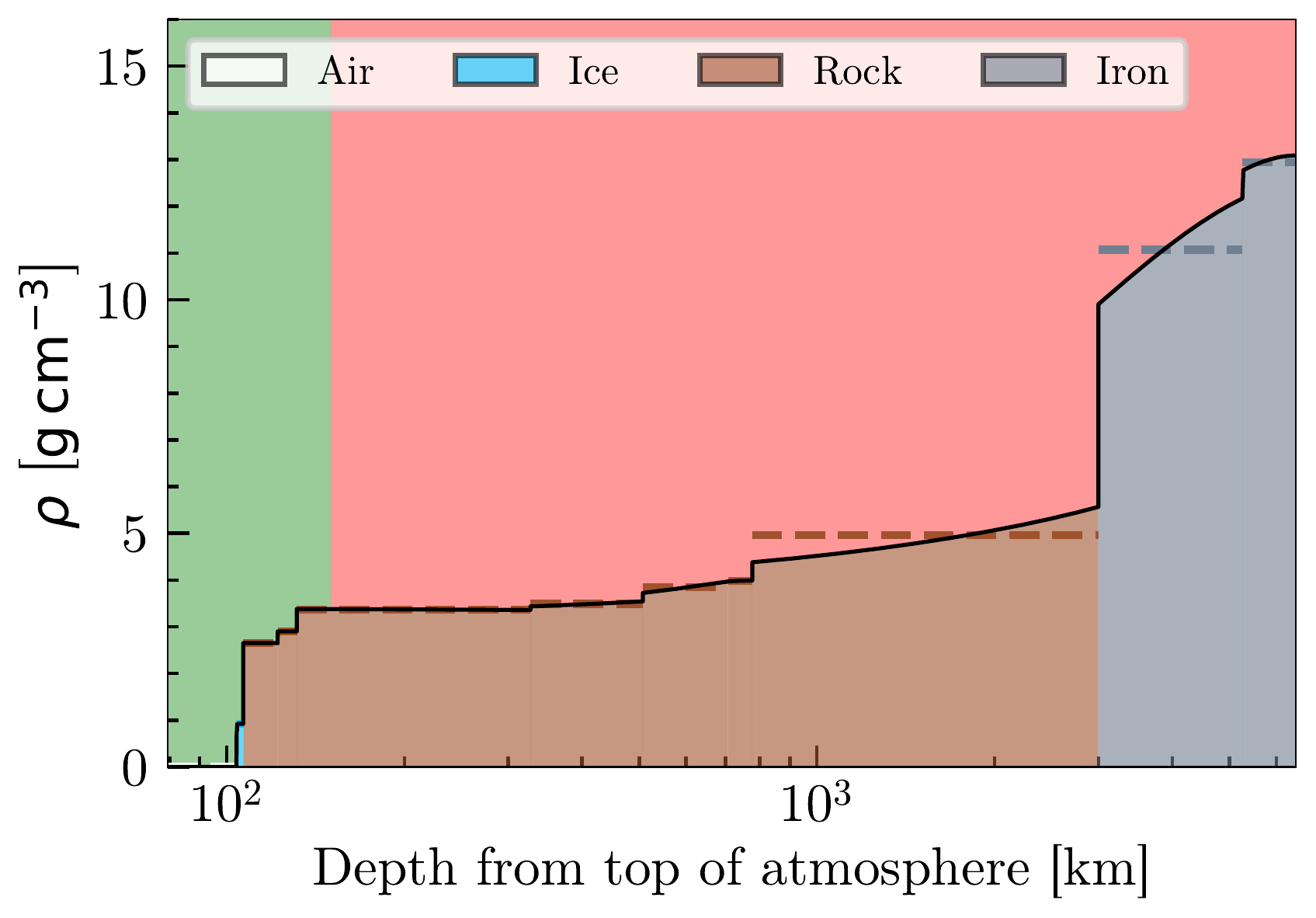}
  \caption{
     \textbf{\textit{Earth model for South Pole}}
  }
  \label{fig:earth_south_pole}
\end{figure}

\subsection{TRIDENT Earth Model}

The water at the proposed TRIDENT deployment site is 3,475~m deep.
Thus the TRIDENT Earth model is the PREM model with 3,475~m of water, and 103~km of air above it.

\begin{figure}[h]
  \centering
  \includegraphics[width=0.6\textwidth]{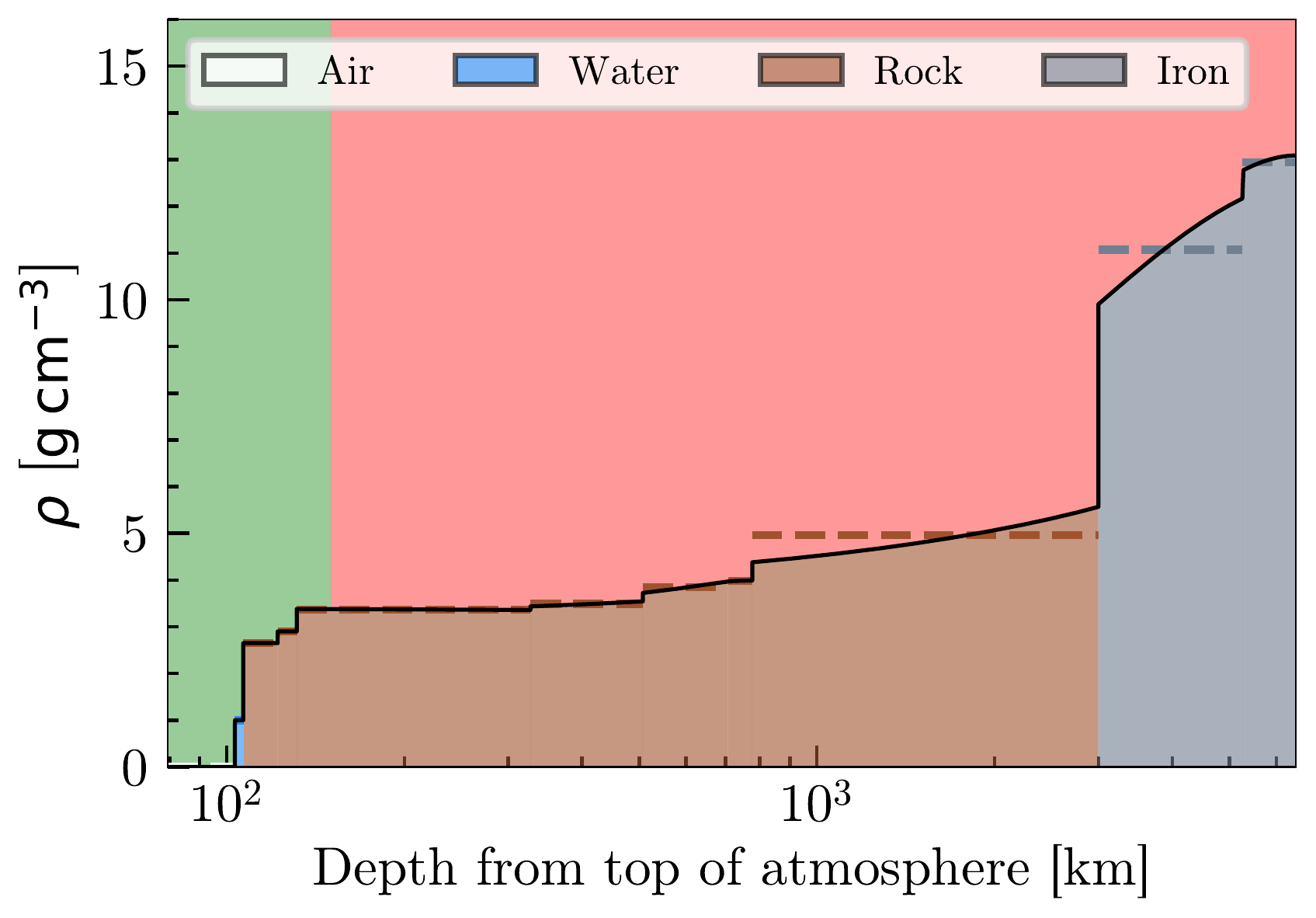}
  \caption{
     \textbf{\textit{Earth model for the TRIDENT detector.}}
  }
  \label{fig:earth_trident}
\end{figure}

\subsection{Water Earth Model}
\begin{figure}[h]
  \centering
  \includegraphics[width=0.6\textwidth]{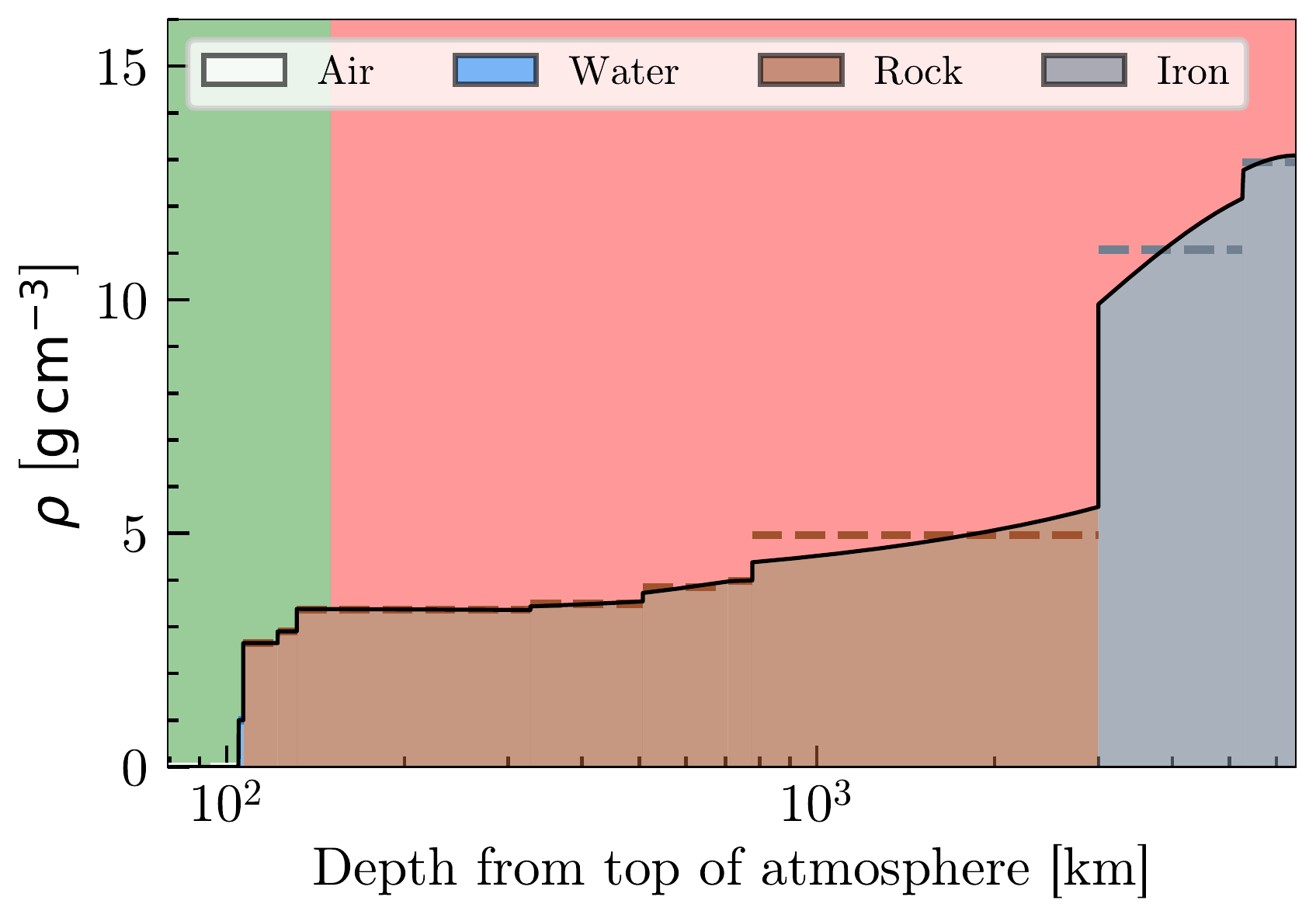}
  \caption{
     \textbf{\textit{Earth model for generic water detector.}}
  }
  \label{fig:earth_water}
\end{figure}